\documentstyle[aps,eqsecnum,preprint,multicol]{revtex}
\input epsf

\begin{document}
\title{\bf Phase behaviour and thermodynamic anomalies of core-softened fluids}
\author{Nigel B. Wilding}
\address{Department of Physics, University of Bath, Bath BA2 7AY, U.K.}
\author{James E. Magee}
\address{Department of Physics and Astronomy, The University of Edinburgh,\\
Edinburgh EH9 3JZ, U.K.}
\tighten
\maketitle

\begin{abstract} 

We report extensive simulation studies of phase behaviour in single
component systems of particles interacting via a core-softened
interparticle potential. Two recently proposed examples of such
potentials are considered; one in which the hard core exhibits a
shoulder, (Sadr-Lahijany {\em et al}, Phys. Rev. Lett. {\bf 81}, 4895
(1998)) and the other in which the softening takes the form of a linear
ramp (Jagla, Phys. Rev. {\bf E63}, 061501 (2001)). Using a combination
of state-of-the-art Monte Carlo methods, we obtain the gas, liquid and
solid phase behaviour of the shoulder model in two dimensions. We then
focus on the thermodynamic anomalies of the liquid phase, namely maxima
in the density and compressibility as a function of temperature.
Analysis of the finite-size behaviour of these maxima suggests that,
rather than stemming from a metastable liquid-liquid critical point, as
previously supposed, they are actually induced by the quasi-continuous
nature of the two dimensional freezing transition. For the ramp model
in three dimensions, we confirm the existence of a stable liquid-liquid
(``second'') critical point occurring at higher pressure and lower
temperature than the liquid-gas critical point. Both these critical
points and portions of their associated coexistence curves are located
to high precision. In contrast to the shoulder model, the observed
thermodynamic anomalies of this model are found to be authentic, i.e.
they are not engendered by an incipient new phase. We trace the locus
of density and compressibility maxima, the former of which appears to
converge on the second critical point.

\end{abstract}
\epsfclipon 

\pacs{PACS numbers: 64.60.Fr, 05.70.Jk, 68.35.Rh, 68.15.+e}


\section{Introduction}

Much attention has been paid recently to the phase behaviour of single
component systems of particles interacting via so-called core-softened
(CS) potentials.  These potentials possess a repulsive core that
exhibits a region of ``softening'' in the form of a shoulder or a ramp.
Physical motivation for such models derives from the desire to
encapsulate within a simple two-body isotropic potential, the 
complicated features of systems interacting via anisotropic potentials.
Notable examples of the latter include liquid metals \cite{CUMMINGS81},
tetrahedrally bonded molecular liquids such as phosporous
\cite{KATAYAMA00}, and water \cite{STILLINGER}. Performing such
simplifications yield models that are analytically and computationally
tractable but which, one hopes, nevertheless retain the qualitative
physical features of the real systems they seek to describe.

Notwithstanding their relevance to real anisotropic systems, model CS
systems have long been studied for their intrinsic physical interest
\cite{HEMMER70,KINCAID76,DEBENEDETTI91,BORICK93,JAGLA98}. Indeed it is
well established, that they exhibit significantly richer phase
behaviour than conventional single component fluids. For instance, in
certain CS models an isostructural solid-solid phase transition is
observed, accompanied by a solid-solid critical point
\cite{KINCAID76,BOLHUIS97}. But perhaps the most intriguing feature of
core-softening is the prediction that it may engender a demixing
transition between two liquids of different densities, distinct and
additional to the usual liquid-gas phase transition. 

The first suggestions along these lines came from the perturbative
calculations of Stell and co-workers \cite{HEMMER70,KINCAID76} more
than 30 years ago. Much more recently, Stanley and co-workers
\cite{SL98,SCALA01,BULDYREV02} have presented simulation evidence
apparently supporting the existence of a liquid-liquid phase transition
in a two-dimensional (2d) system of CS particles. These studies show
that on reducing the temperature at constant pressure, a maximum occurs
in the density, whilst the compressibility passes through a minimum
before subsequently rising strongly as the system approaches the
freezing transition. These thermodynamic anomalies,
specifically the rise in the compressibility with decreasing
temperature, were attributed to the existence of a liquid-liquid
critical point (termed the ``second critical point''). No direct
evidence for this critical point was found in the stable liquid region.
However, a power law extrapolation of the measured increase of the
compressibility as a function of temperature, suggested that a critical
point may lie hidden within the stable crystalline region. This
proposal, that the thermodynamic anomalies are linked to a metastable
second critical point, was reinforced by mean field calculations, based
on a simple cell model. These were reported to indicate a critical
point whose location was consistent with that found from extrapolation
of the measured rise in the compressibility.

The discovery of thermodynamic anomalies in CS fluids  mirrors similar
findings in liquid water \cite{SCIORTINO97,SPEEDY76} close to the
freezing transition. The apparent connection with a second critical
point lends weight to the hypothesis that a metastable liquid-liquid
critical point may be responsible for the celebrated anomalous
behaviour of water \cite{SCIORTINO97,STANLEY94,STANLEY98,SCALA00}.
Although no compelling evidence for a liquid-liquid phase transition in
supercooled water has yet been reported, there does exist a variety of
indirect experimental and theoretical data favouring the proposition
\cite{BRAZHKIN97,MISHIMA00,SOPER00}. Additionally, there is evidence 
for liquid-liquid phase transitions in a number of other single
component systems such as liquid molecular phosporous
\cite{KATAYAMA00}, graphite \cite{TOGAYA97}, silica
\cite{LACKS00,VOIVOD01}, as well as certain molecular models that take
directional bonding into account \cite{ROBERTS96,VEGA98}.

To date, most studies of second critical points in CS systems appear to
indicate a liquid-liquid transition that is {\em metastable} with
respect to crystallization \cite{BULDYREV02,MALESCIO02,JAGLA99}. By
contrast, Jagla \cite{JAGLA01} has recently presented Monte Carlo (MC)
simulation results for a CS model which, he submits, provides evidence
of a {\em stable} liquid-liquid critical point. This model differs from
many other CS models (see eg. ref.~\cite{SL98}) in that rather than
having a pronounced shoulder in the potential, the softening takes the
form of an inclined linear ramp. Studies of the model found
van-der-Waals type loops along isobars and evidence of a liquid-phase
density anomaly.

Inspired by the above findings, we have attempted to gather further
simulation evidence for the existence of liquid-liquid critical points
in core softened models. Our approach employs a variety of Monte Carlo
simulation methods tailored to the efficient study of liquid and solid
phases and their critical points. We investigate two CS models, 
qualitatively distinct in character. The first is the ``shoulder''
potential initially proposed by Sadr-Lahijany {\em et al} \cite{SL98},
which we study in 2d. The other is  Jagla's ``ramp'' potential
\cite{JAGLA01} which we study in 3d. 

The main features of our results are as follows. In the 2d shoulder
model we reproduce the phase diagram found in ref. \cite{SL98} and
present new results concerning the solid-solid phase transitions. We
then proceed to a detailed study of the liquid phase, focusing
attention on the  density and compressibility anomalies and their
finite-size behaviour. Our results show that rather than being linked
to a metastable liquid-liquid critical point, these anomalies are
instead associated with the freezing of the liquid to a 2d solid of
lower density via a quasi-continuous phase transition.  We also report
cell model mean-field calculations for the shoulder model. These,
however, do not support the existence of a liquid-liquid critical point reported on
the basis of similar calculations in ref.\cite{SL98}.

Our studies of the 3d ramp model deploy isothermal-isobaric MC
simulation methods to study the reported finding of a stable second
critical point \cite{JAGLA01}. We indeed confirm the existence of this
point, and determine its parameters to higher precision that obtained
previously. Using multicanonical MC sampling and histogram reweighting
techniques, we map the liquid-liquid and liquid-gas coexistence lines
to high precision. Investigation of the low density liquid phase
confirms the existence of maxima in the density and compressibility. In
contrast to the shoulder model, these anomalies are found to be
authentic i.e. they are not the result of an incipient new phase. We
trace the locus of density and compressibility maxima and find that the
former appears to converge on the liquid-liquid critical point, while
the latter intersect the phase boundary somewhat below the critical
temperature.

\section{Shoulder model}

The first CS model we consider takes the form of a $12$-$6$ Lennard
Jones (LJ) potential whose attractive tail is modified by the addition
of a Gaussian well centered on radius $r=r_0\:$. 

\begin{equation}
u(r)=4\epsilon\left[\left(\frac{\sigma}{r}\right)^{12} 
-\left(\frac{\sigma}{r}\right)^6\right]-
\epsilon\lambda\exp\left[-\left(\frac{w(r-r_0)}{\sigma}\right)^2\right]
\label{eq:slpot}
\end{equation}

\begin{figure}[h]
\setlength{\epsfxsize}{16.0cm}
\centerline{\epsffile{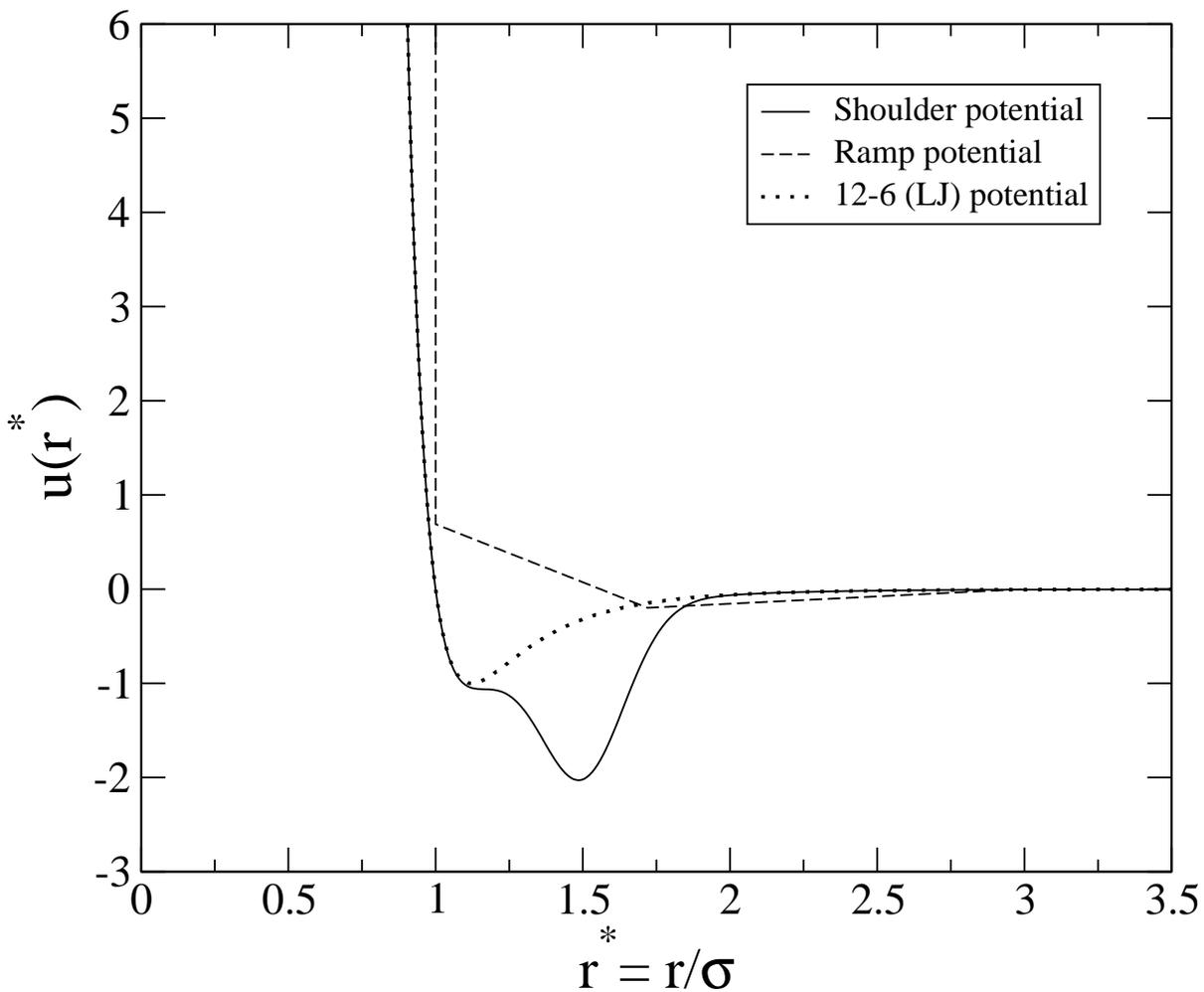}}
\vspace*{5mm}
\caption{The core-softened potentials studied in this work. The
shoulder potential of ref.~\protect\cite{SL98} is represented by the solid
curve; the dashed curve corresponds to the ramp potential of
ref.~\protect\cite{JAGLA01}. Also shown for comparison (dotted curve) is the
standard Lennard-Jones 12-6 potential.}

\label{fig:pots}
\end{figure}

In common with ref. \cite{SL98}, we have studied this model in
{\em two-dimensions} employing potential parameters values $\epsilon=1,
\lambda=1.7, \sigma=1, w=5, r_0=1.5$.  Additionally, a cutoff was
imposed at $r=2.5\sigma$ for which no correction was applied. The form
of the resulting potential is shown in fig.~\ref{fig:pots}. From
the figure, one see that in the core and tail regions, the potential closely
approximates the LJ form. Close to $r_0$, however, the potential is dominated by
the inverted Gaussian, the effect of which is to generate a shoulder in
the core.

We have studied the gas, liquid and solid phases of this model using 
MC simulation techniques. In the following subsections we describe our
results for each region of the phase diagram in turn.

\subsection{The liquid-gas and sublimation lines}
\label{sec:sublimation}

In terms of computational tractability, the liquid-gas transition line
is the most straightforwardly obtained feature of the phase diagram
because it lies in a region of relatively elevated temperature, where
acceptance rates for Monte Carlo moves are generally high.
Experience with a variety of fluid models has shown
\cite{WILDING95,PANAGIO00,WILDING01} that a combination of grand
canonical ensemble (GCE) simulations \cite{FRENKEL}, multicanonical
preweighting \cite{BERG92} and histogram reweighting \cite{FERRENBERG}
provides an efficient means of tracing a liquid-gas coexistence curve.
Details of the implementation can be found in
refs.~\cite{WILDING95,WILDING01}, but the basic strategy is to focus on
the fluctuations of the number density  $\rho=N/L^d$; more specifically,
the number density probability distribution function $p(\rho)$.
Precisely at coexistence, this distribution is doubly peaked with equal
integrated weight (area) under each peak \cite{BORGS92}. For each
chosen value of the reduced temperature $T^\star=k_BT/\epsilon$,
coexistence is located by tuning $\mu^\star\equiv\mu/k_BT$ until the
measured $p(\rho)$ satisfies this equal peak weight criterion. 

\begin{figure}[h]
\setlength{\epsfxsize}{8.5cm}
\centerline{\epsffile{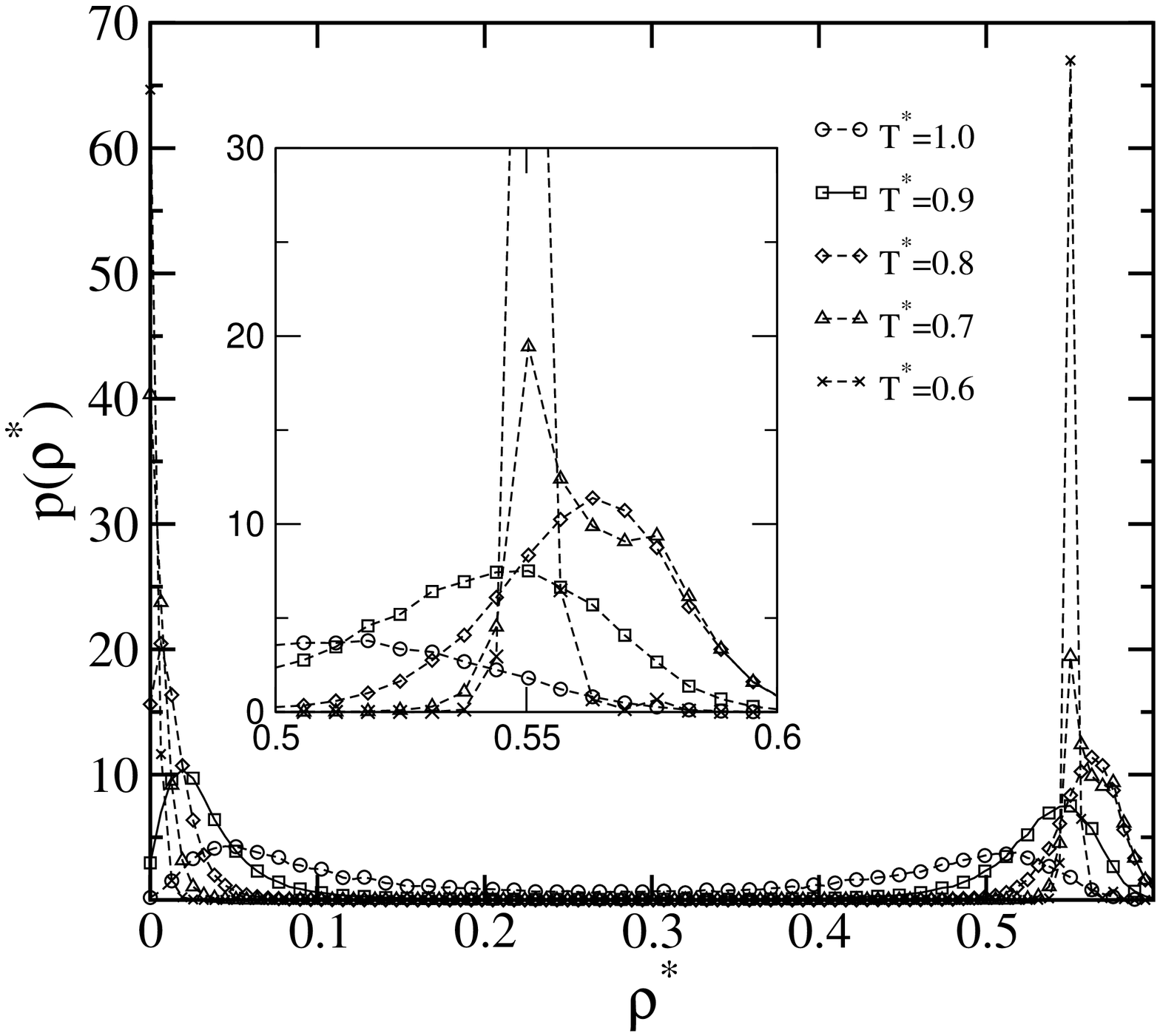}}
\setlength{\epsfxsize}{8.5cm}
\centerline{\epsffile{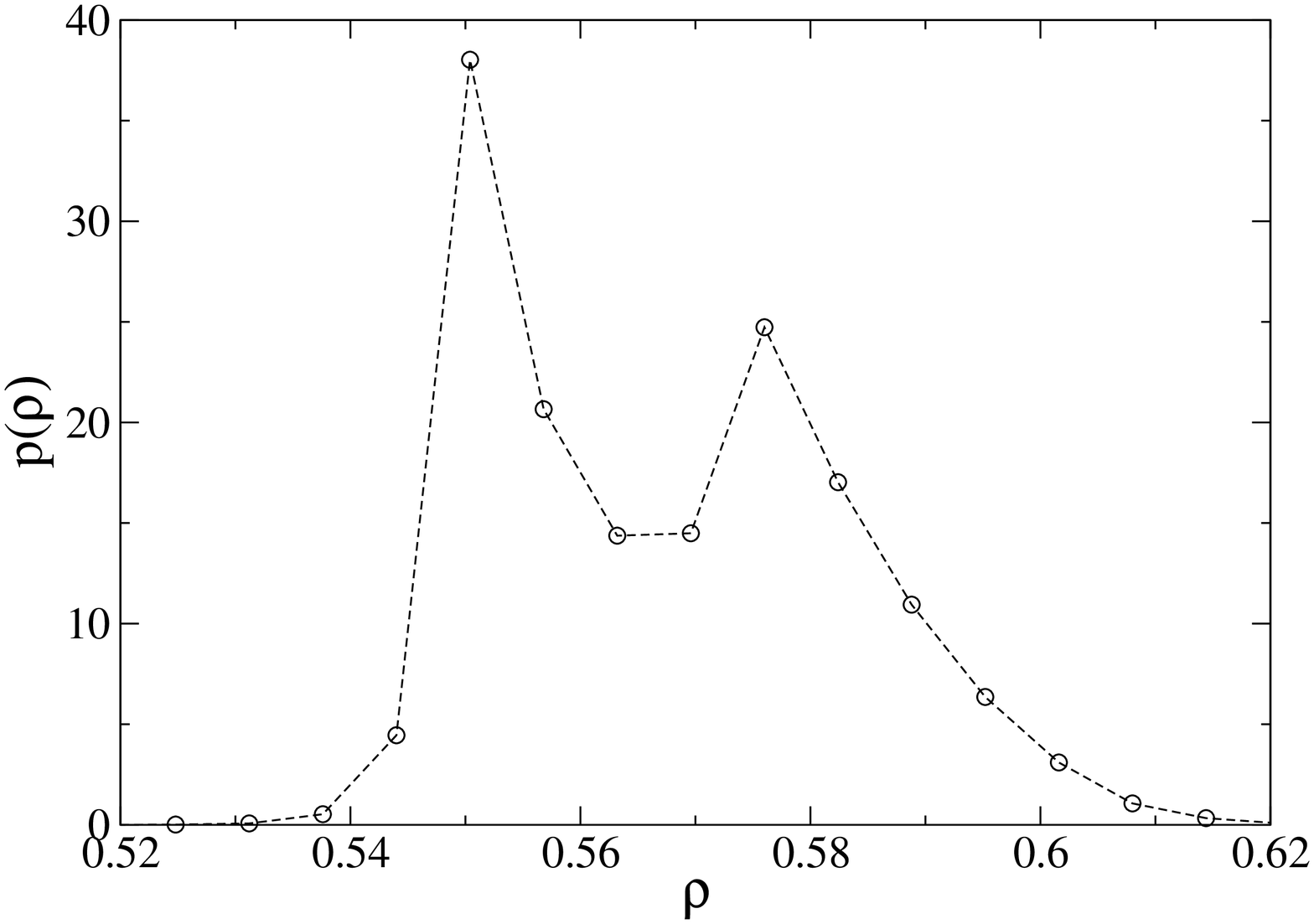}}
\vspace*{5mm}

\caption{{\bf(a)} The measured density distribution $p(\rho)$ at a
selection of state points along the liquid-gas and LDTS-gas coexistence
lines of the shoulder model, obtained in the manner described in the text. The inset shows
a magnified version of the high density region. Dashed lines serve as
guides to the eye. {\bf (b)} The double peaked form of $p(\rho)$ at
$T^\star=0.68$ corresponding to LDTS-liquid coexistence, obtained by
histogram extrapolation of the triple point distribution at
$T^\star=0.72$ shown in (a).}

\label{fig:condists}
\end{figure}

We have measured the form $p(\rho)$ along the liquid-gas coexistence
curve, starting near the critical point. The simulations were performed
for systems of linear dimension $L=10\sigma$ and $L=12.5\sigma$. A
selection of coexistence distributions for the larger system size are
presented in fig.~\ref{fig:condists}(a). In the vicinity of the
liquid-gas critical point the two peaks are quite close together and
the trough separating them is shallow. In principle, the critical
temperature can be estimated precisely by employing finite-size scaling
(FSS) methods according to the approach described in
\cite{BRUCE92,WILDING95}. However, as the critical behaviour is not the
principal focus of the present study we have not performed a full FSS
analysis. Instead we obtained an approximate estimate of the critical
parameters by tuning $T^\star$ and $\mu^\star$ until $p(\rho)$ matches
the known universal fixed point form of the order parameter
distribution function appropriate to 2d Ising universality class
\cite{BRUCE92}. This yielded the estimate $T^\star_c=0.955(5),
\mu_c^\star=-3.373(2)$.

As the temperature is decreased below its critical value, the trough
separating the peaks deepens and the peak densities separate in the
usual way \cite{BRUCE92}. On reaching $T^\star=0.72$, however, an
interesting feature emerges--the liquid peak bifurcates, with a new
narrow peak appearing on its low density side (fig.~\ref{fig:condists}(a)).  On further reducing
temperature, this new peak grows at the expense of the original peak.
Below $T^\star=0.6$ only the new peak remains. 

\begin{figure}[h]
\setlength{\epsfxsize}{10.0cm}
\centerline{\epsffile{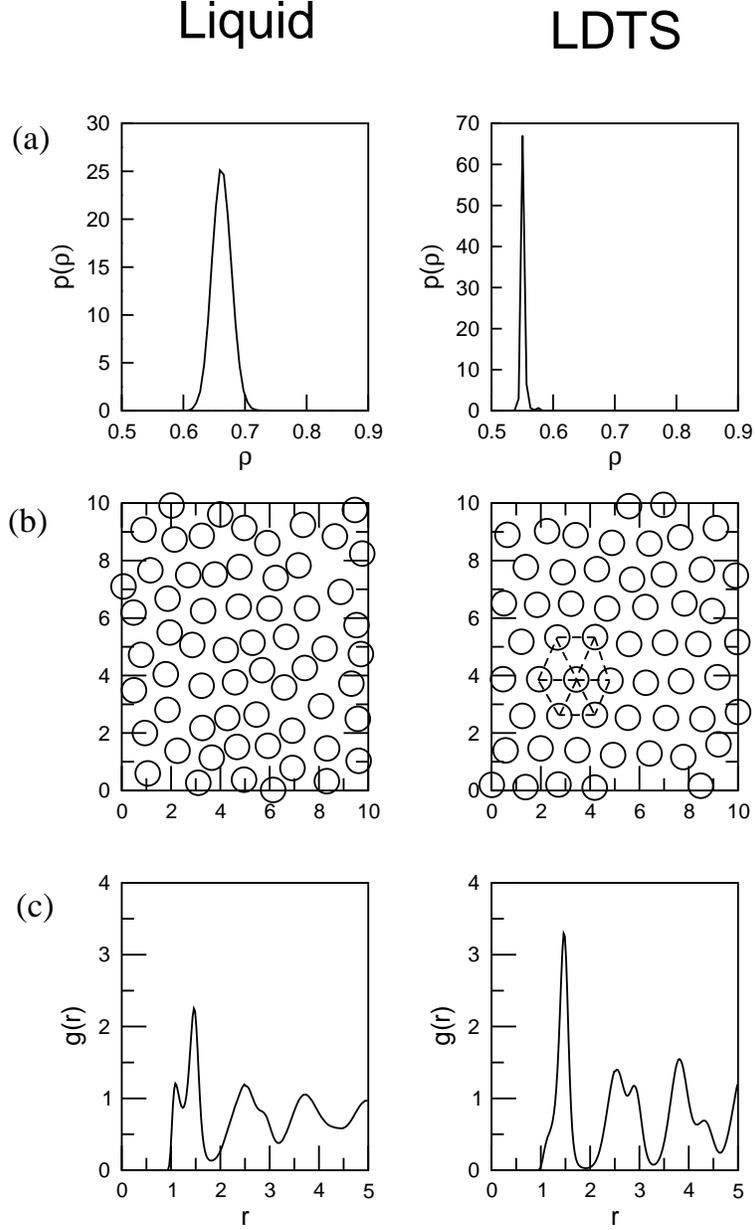}} 
\vspace*{5mm}
\caption{{\bf (a)} Density distributions $p(\rho)$ for the liquid at
$T^\star=0.75, \mu^\star=-3.7$ (left) and the LDTS at $T^\star=0.65,
\mu^\star=-3.8$ (right) for a system of size $L=10\sigma$. {\bf (b)}
Snapshot configurations of the liquid and LDTS. {\bf (c)} Measured
radial distribution function $g(r)$ for the liquid (left) and LDTS
(right).}
\label{fig:lds_lconf}
\end{figure}

The appearance of three peaks in $p(\rho)$ at $T^\star=0.72$ is
suggestive of a triple point. In order to identify the phases involved,
we have examined the structure of configurations having densities close
to those of the peaks. The lowest density phase, having $\rho< 0.02$ is
immediately identifiable as a gas phase. In order to isolate the two
high density phases, we studied temperatures slightly above and below
$T^\star=0.72$.  Snapshot configurations and measurements of the radial
distribution function $g(r)$ for the $L=10\sigma$ system
(fig.~\ref{fig:lds_lconf}(a)) show that above $T^\star=0.72$ the system
is liquidlike, having a nearest neighbor distance corresponding to the
radius of the shoulder in the potential. Below $T^\star=0.72$, the high
density phase (which has a density {\em less} than that of the liquid,
see fig.~\ref{fig:lds_lconf}(a)) is a 2d solid of triangular symmetry,
with a nearest neighbor distance consistent with the radius of the
potential minimum $r_0$ (c.f. fig.~\ref{fig:pots}). This is the low
density triangular solid (LDTS) identified in ref.~\cite{SL98}.

\begin{figure}[h]
\setlength{\epsfxsize}{9.0cm}
\vspace*{1cm}
\centerline{\epsffile{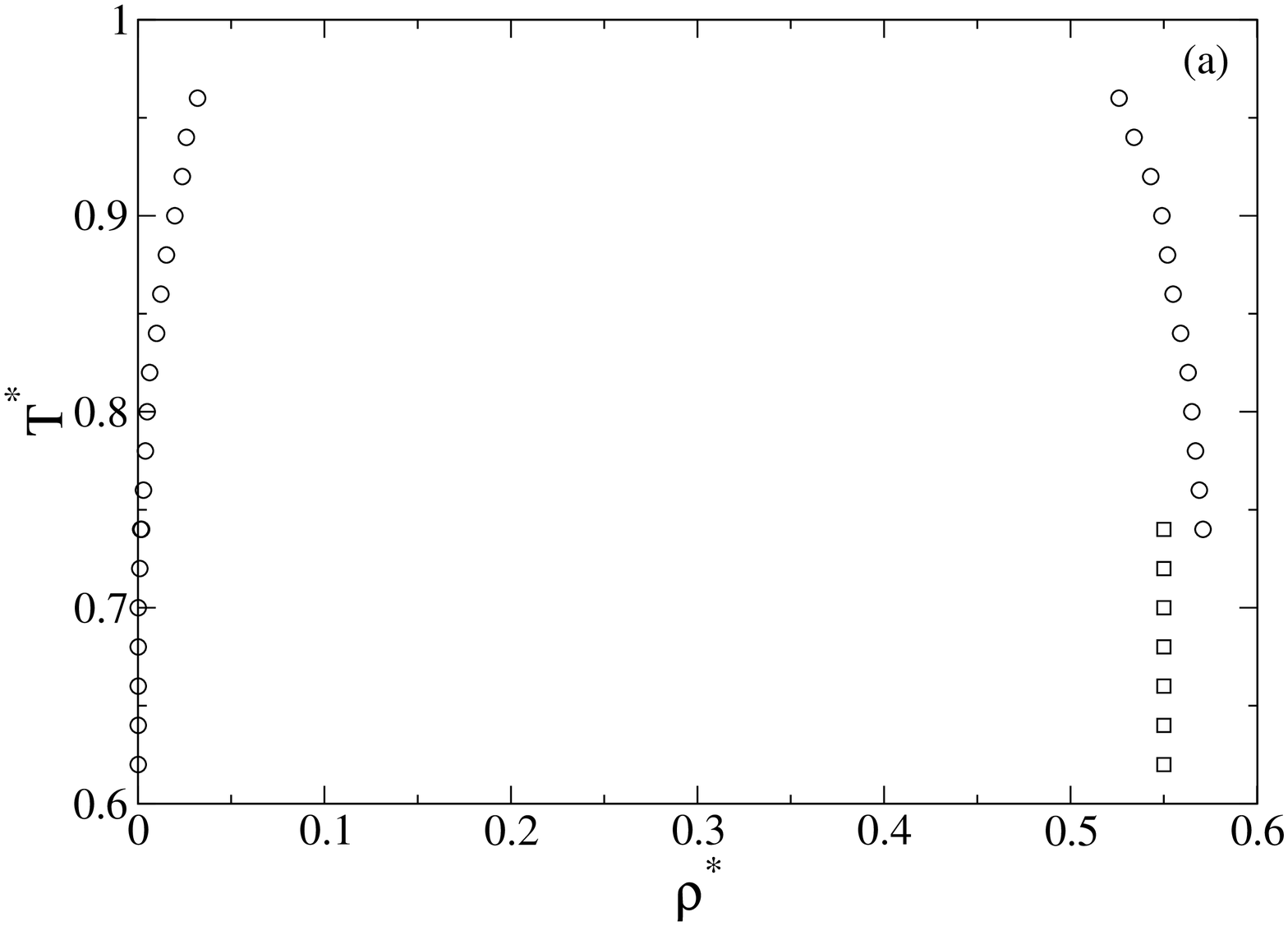}} 
\setlength{\epsfxsize}{9.0cm}
\centerline{\epsffile{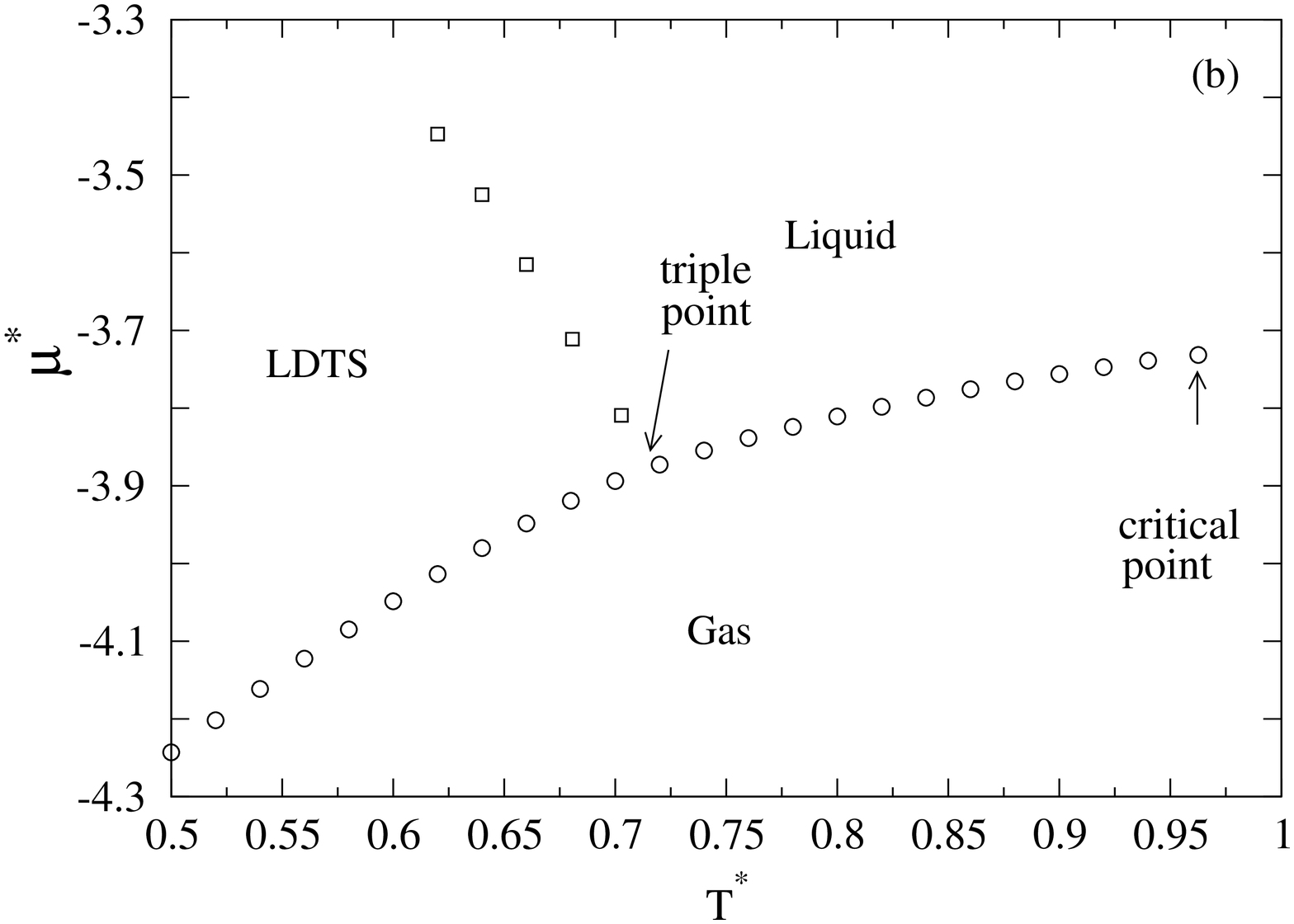}} 
\vspace*{5mm}

\caption{{\bf (a)} The liquid-gas and LDTS-gas coexistence lines in 
the $\rho-T^\star$ plane, obtained as described in the text. {\bf (b)}
The corresponding phase diagram in the $\mu^\star-T^\star$ plane.  Also
shown is a segment of the finite-size shifted LDTS-liquid coexistence
line obtained for $L=12\sigma$ via reweighting of the triple point
histograms (cf. fig.\protect\ref{fig:condists}(b))}

\label{fig:gaspds}
\end{figure}

Generally speaking, GCE Monte Carlo is unable to deal well with solid
phases due to the very low acceptance rate for particle exchange
moves at typical solid densities. In the present case, however, it
turned out to be possible to follow the sublimation line to
temperatures considerably below that of the triple point because of the
unusually low density of the LDTS. Our results
(fig.~\ref{fig:gaspds}(a)) show that the density of the LDTS remains
constant within the resolution of our density scale. This finding would
appear to reflect the fact that the functional form of the potential
minimum in which the particles reside, is approximately {\em symmetric}
(it is Gaussian), implying that there is no energetic advantage of
contraction over expansion.

Figs.~\ref{fig:gaspds}(b) show the liquid-gas and sublimation lines in
the space of $\mu^\star$--$T^\star$.  Overall, little system size
dependence was observed in the locus of the liquid-gas and LDTS-gas
coexistence lines. A notable exception was the neighborhood of the
triple point at which the LDTS, liquid and the gas all coexist. This
point marks the start of a coexistence line between the LDTS and the
liquid. Because the solid is less dense than the liquid with which it
coexists, the gradient $d\mu^\star/dT^\star$ of this coexistence line
is negative, as mandated by the Clausius-Clapeyron equation. A portion
of the LDTS-liquid coexistence line was estimated by histogram
reweighting the triple point data for the $L=12\sigma$ system size. The
results, (included in fig.~\ref{fig:gaspds}(b)) were obtained by tuning
$T^\star$ and $\mu^\star$ such as to maintain both a liquid and a solid
peak in $p(\rho)$ (cf. fig.~\ref{fig:condists}(b)) for an example
distribution) . It was subsequently found, however, that the locus of
the LDTS-liquid line exhibits large finite-size effects; specifically
it shifts to lower temperature with increasing system size. We shall
address this matter in detail in sec.~\ref{sec:freezing}, where we also
present a more accurate estimate for the LDTS-liquid coexistence
boundary. 

\subsection{Liquid-solid transitions and thermodynamic anomalies}
\label{sec:freezing}

In this section we consider the properties of the liquid phase and its
freezing behaviour. These have been investigated by means of GCE
simulations conducted along two lines of constant chemical potential,
one having $\mu^\star=-3.0$ and the other having $\mu^\star=-3.5$.
Simulation runs were performed at a number of selected temperatures
along both iso-$\mu^\star$ lines, extending from the freezing point up
to $T^\star=0.9$. The principal observable recorded in each run was the
instantaneous density, the fluctuation spectrum of which permits
construction of the density distribution $p(\rho)$ (accumulated as a
histogram) and thence evaluation of the average density
$\langle\rho\rangle$ and the compressibility $\kappa$. 

The data from the complete set of runs performed along each
iso-$\mu^\star$ line, were combined self consistently within the
multihistogram framework \cite{FERRENBERG}.  Histogram reweighting was
then employed to accurately interpolate into the regions of temperature
intermediate between the simulation state points.  In order to allow
assessment of finite-size effects in the results, this procedure was
repeated for five system sizes: $L=17.5\sigma, 22.5\sigma, 30\sigma,
35\sigma$ and $L=40\sigma$.  For the largest system sizes, reliable
histogram extrapolation necessitated runs at ten different temperatures
along an iso-$\mu^\star$ line. For the smallest system sizes, five
simulations proved sufficient.

\begin{figure}[h]
\setlength{\epsfxsize}{9.0cm}
\centerline{\epsffile{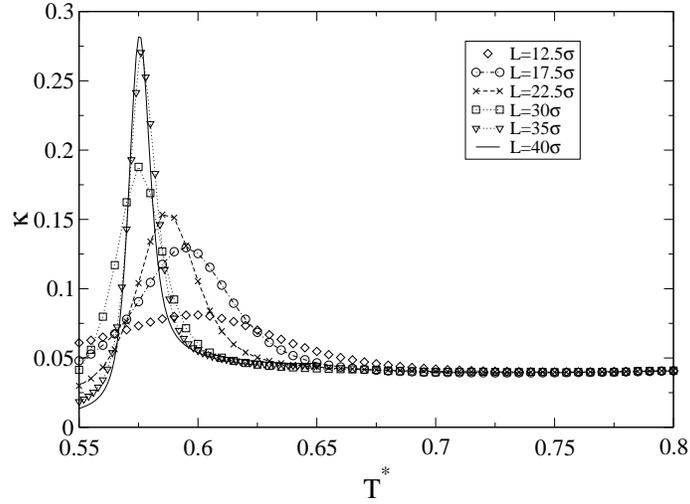}} 

\vspace*{5mm}

\caption{The measured temperature dependence of the compressibility
$\kappa=\beta V\langle(\Delta\rho)^2\rangle$ for $\mu^\star=-3.0$ for the system
sizes shown in the key. The curves were obtained from multi-histogram
extrapolation of the data from a number of individual simulations, as
described in the text.}

\label{fig:mu3.0}
\end{figure}

Fig.~\ref{fig:mu3.0} shows the measured finite-size dependence of the
compressibility for $\mu^\star=-3.0$. One observes that on reducing the
temperature, the compressibility passes through a shallow minimum at
$T^\star\approx 0.75$, before rising strongly to a peak. The position
of this peak is, for the smaller system sizes, strongly finite-size
dependent, moving both to lower temperature and becoming considerably
sharper as $L$ increases. For the two largest system sizes
($L=35\sigma$ and $L=40\sigma$) there is, however, relatively little
difference in the peak height or position, which lies at
$T^\star=0.576(1)$. We postpone detailed discussion of this finite-size
behaviour until sec.~\ref{sec:concs}.

\begin{figure}[h]
\setlength{\epsfxsize}{9.0cm}
\centerline{\epsffile{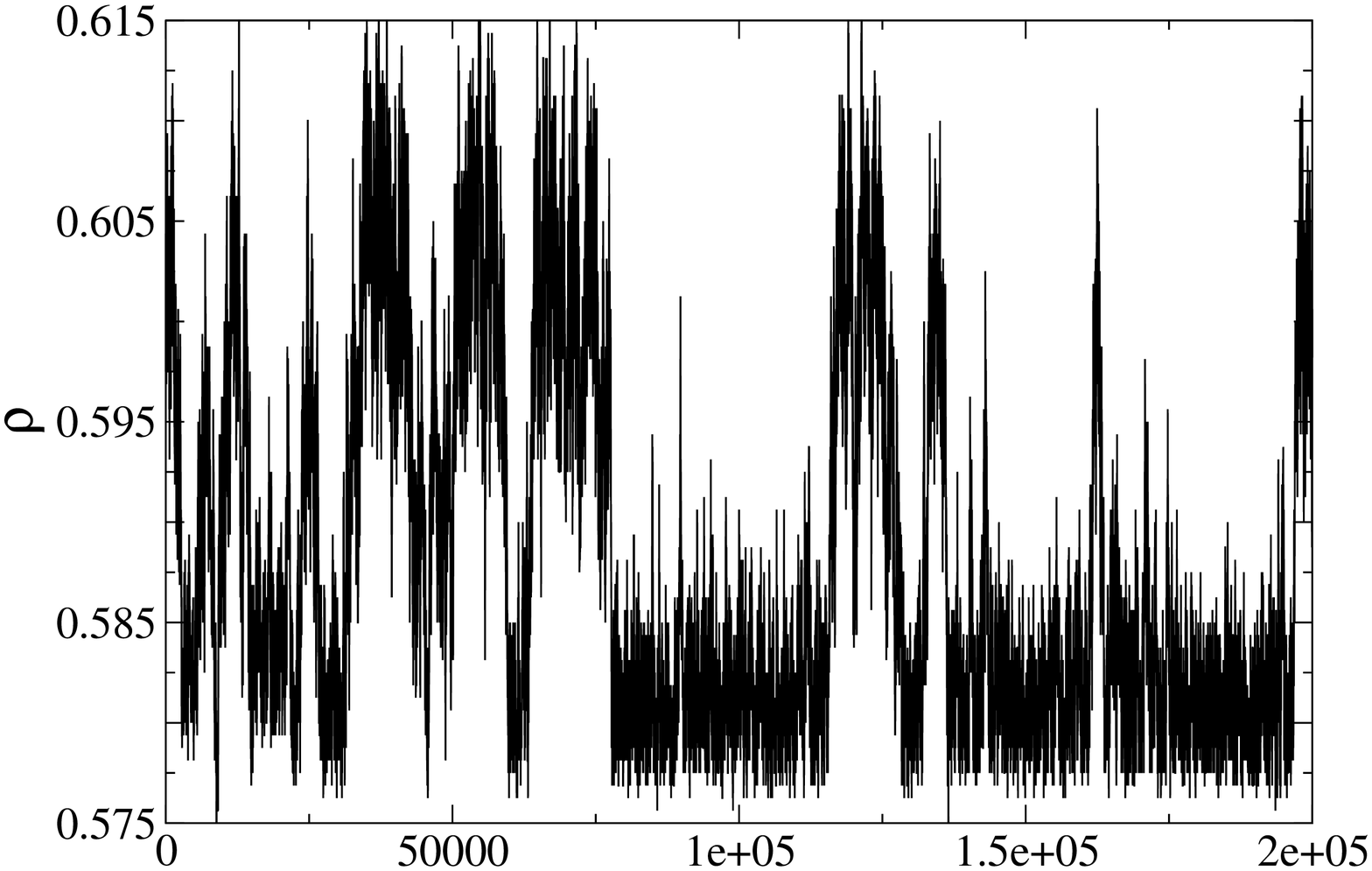}} 
\vspace*{1cm}
\setlength{\epsfxsize}{9.0cm}
\centerline{\epsffile{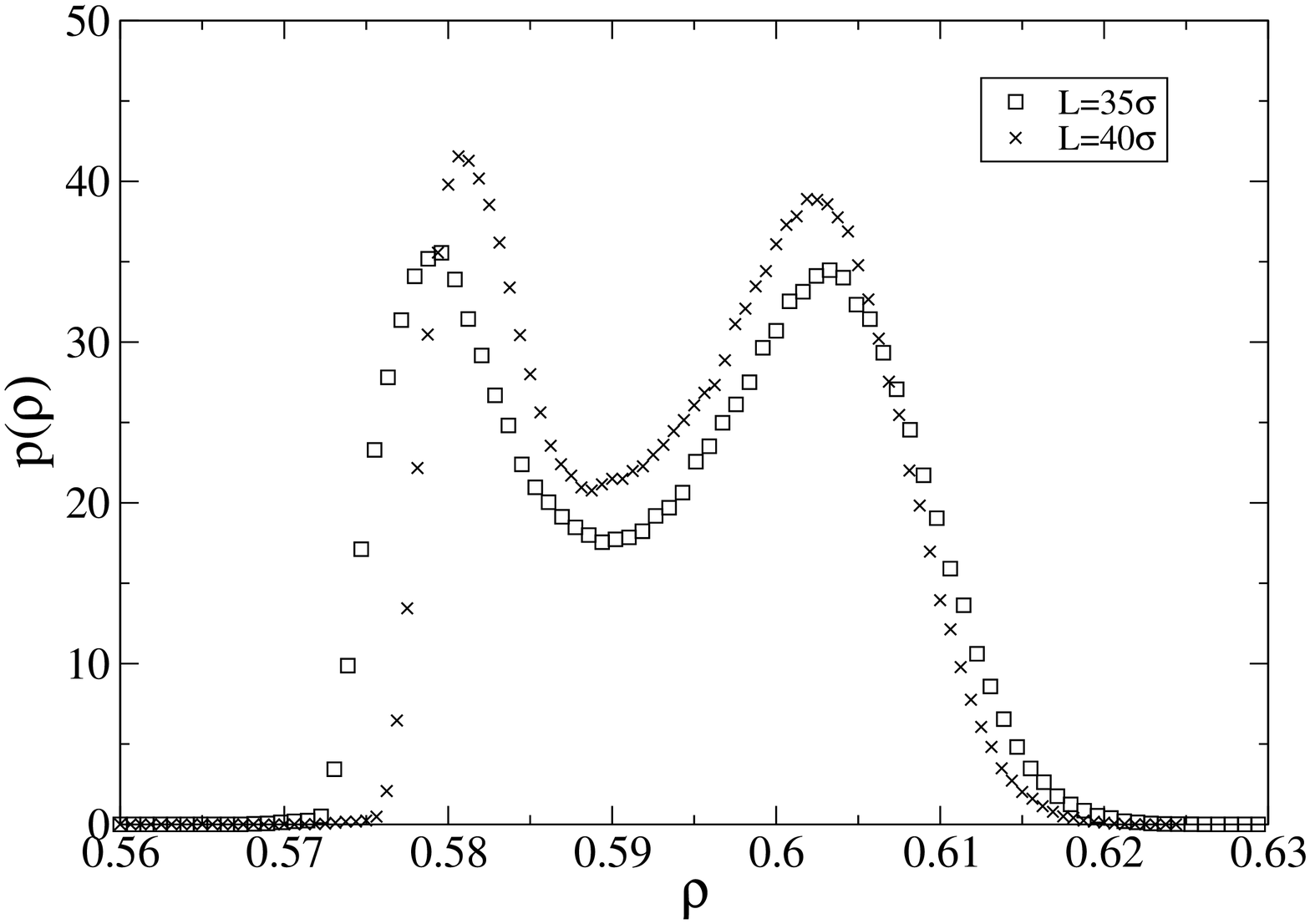}} 
\vspace*{5mm}
\caption{{\bf (a)} A portion of the time evolution of the density for
$L=40\sigma$ at $T^\star=0.576$.{\bf (b)} Density distributions
$p(\rho)$ for $L=35\sigma, 40\sigma$ at the temperature of the compressibility maximum.}
\label{fig:distsmu3.0}
\end{figure}

The increase in the compressibility is traceable to a widening of
$p(\rho)$.  In order to clarify the physical origin of this effect, it
is instructive to examine the time evolution of the system density for
the largest system size $L=40\sigma$ at the temperature for which the
peak occurs. A representative portion of this evolution is shown in
fig.~\ref{fig:distsmu3.0}(a). Clearly there are two regions of
preferred density, and a substantial density autocorrelation time
associated with each. Fig.~\ref{fig:mu3.0}(b) displays the density
distribution $p(\rho)$, for the largest two system size. These exhibit
a double peaked structure. The increase in the compressibility reflects
the widening of $p(\rho)$ as the double peaked structure develops. 

The double peaked form of $p(\rho)$ suggests that two distinct phases
coexist at the temperature of the compressibility peak. To investigate
this matter further, we have studied the configurational structure
at temperatures slightly below and above the peak temperature.
Representative snapshots are shown in fig.~\ref{fig:freezeconfs}. At a
temperature, $T^\star=0.55$, below the transition, 
fig.~\ref{fig:freezeconfs}(a) shows the presence of long ranged
translational order (on the scale of our system size) in the form of a
triangular lattice. This we associate with the LDTS. At $T^\star=0.6$,
above the temperature of the compressibility maximum,
fig.~\ref{fig:freezeconfs}(b) shows that the system is clearly
non-crystalline. There is, nevertheless, evidence for substantial
clusters of LDTS-like structure, one of which we have indicated in the
figure.

We have also obtained the temperature-dependence of the average
density, the finite-size behaviour of which is shown in
fig.~\ref{fig:den_mu3.0}(a). One sees that as temperature is decreased,
the density rises to a peak, before falling rapidly to a lower value.
The peak temperature initially shifts to lower values as the system size
increases, but seems to converge for the largest systems to a value
higher than that to which the compressibility maximum converges. We
find that the temperature at which the density starts to fall rapidly,
coincides with the first appearance of a subsidiary peak in the density
distribution $P(\rho)$ for all systems sizes (cf.
fig.~\ref{fig:den_mu3.0}(b)). This peak occurs at the same density as
the lower density (LDTS) peak in fig.~\ref{fig:distsmu3.0}(b) and is thus
indicative of infrequent fluctuations of the system into the solid
phase.

\begin{figure}[h]
\setlength{\epsfxsize}{9.0cm}
\centerline{\epsffile{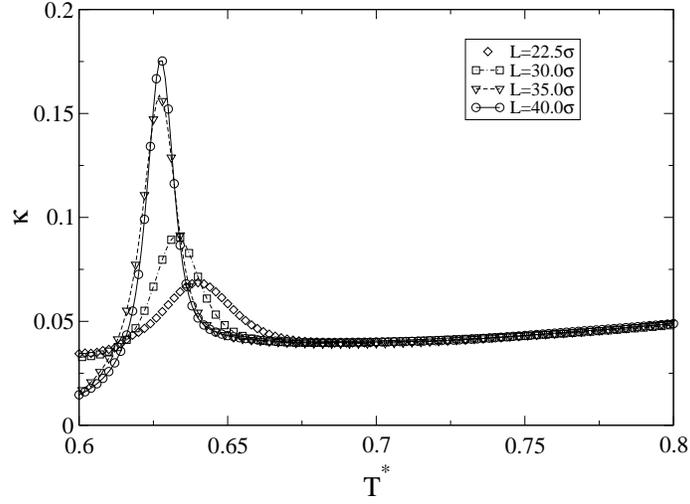}} 
\vspace*{5mm}
\caption{As fig.~\protect\ref{fig:mu3.0}, but for $\mu^\star=-3.5$.}
\label{fig:mu3.5}
\end{figure}

Taken together, the above findings suggest that the thermodynamic
anomalies are tied to the liquid-LDTS freezing line and not a buried
critical point. If so, then similar behaviour should be expected all
along the freezing line. That this is indeed the case is confirmed by
our results for $\mu^\star=-3.5$. This line intersects the LDTS melting
curve close to the gas-liquid-LDTS triple point at a temperature of
$T^\star=0.628(2)$. The associated measurements of the compressibility
(fig.~\ref{fig:mu3.5}) display finite-size effects whose magnitude is
comparable, if not greater than, those observed for $\mu^\star=-3.0$.
The main difference is that for a given system size, the height of the
maximum in the compressibility is some $30\%$ less than at
$\mu^\star=-3.0$. 

\begin{figure}[h]
\setlength{\epsfxsize}{10.0cm}
\centerline{\epsffile{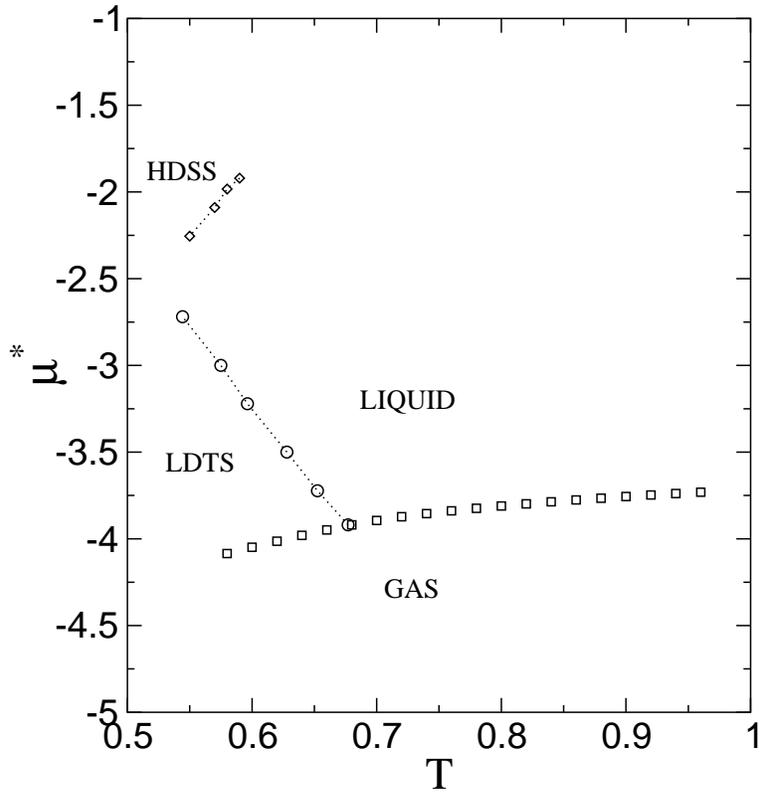}} 
\vspace*{5mm}
\caption{The phase diagram of the 2d shoulder model in the
$\mu^\star-T^\star$ plane, obtained in the manner described in the text. No
analysis of finite-size effects has been performed for the HDSS melting line.}
\label{fig:TmuPD}
\end{figure}

Having located two coexistence state point on the LDTS melting line,  a
number of further simulations were performed (for the $L=40\sigma$
system size) to trace out the whole coexistence line. This was
identified as the locus of compressibility maxima.  The choice of state
points for these runs was guided by histogram extrapolation of the
existing coexistence data. For each $\mu^\star$ studied, visual checks
were made of configurations at temperatures either side of the
compressibility maximum in order to confirm that the line of maxima
coincides with the freezing line. The results are shown in
fig.~\ref{fig:TmuPD}, together with the liquid-gas and sublimation
lines determined in sec.~\ref{sec:sublimation}. 

\begin{figure}[h]
\setlength{\epsfxsize}{6.0cm}
\centerline{\epsffile{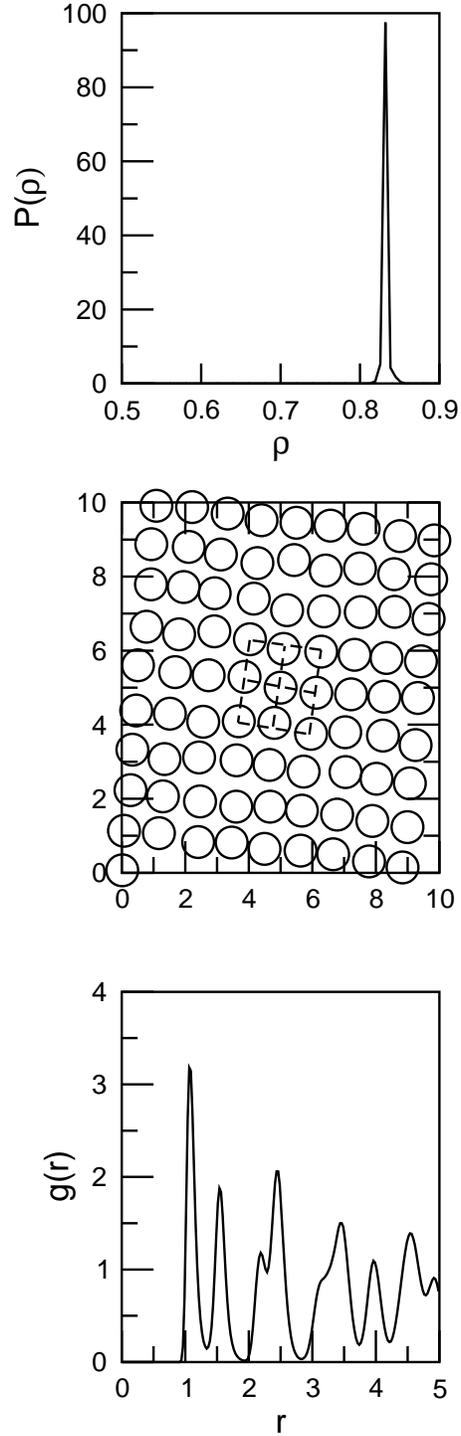}} 
\vspace*{5mm}
\caption{{\bf (a)} Density distribution $p(\rho)$ for the HDSS at
$T^\star=0.6, \mu^\star=-2.0$ for a system of size $L=10\sigma$. {\bf (b)}
A typical snapshot configuration. {\bf (c)} Measured
radial distribution function $g(r)$. }
\label{fig:hds_conf}
\end{figure}

Fig.~\ref{fig:TmuPD} shows that the LDTS melting line is bounded by two
triple points: the gas-LDTS-liquid point at low $\mu^\star$ end, and a
triple point involving the liquid, LDTS and a high density square solid
(HDSS) phase at high $\mu^\star$. The HDSS phase was first reported in
ref.~\cite{SL98}. For values of $\mu^\star$ exceeding that of the
latter triple point, the liquid freezes into the HDSS rather than the
LDTS. We attempted to map the liquid-HDSS coexistence curve.
Unfortunately, at the densities prevailing near this line, our GCE
algorithm becomes very inefficient due to a low acceptance rate for
particle transfers. This prevented us from studying large system sizes
and from determining the finite-size behaviour of the transition. We
have therefore only be able to determine a limited portion of its locus
for one small system of size $L=10\sigma$. The results
(fig.~\ref{fig:hds_conf}), confirm that the HDSS has a square lattice
exhibiting quasi long ranged order. The associated estimate of the
freezing line for this system size is included in fig.~\ref{fig:TmuPD}.

\subsection{Cell theory calculations}
\label{sec:cell}

Ref.~\cite{SL98} offers direct evidence for the existence of a second
critical point in the 2d shoulder model in the form of a mean-field
cell theory calculation. This was reported to show a liquid-liquid
critical point whose position was consistent with the extrapolation
into the stable crystalline region of the simulation measurements of
the compressibility.  In this section we detail our attempts to follow
up this finding.

Cell theory was first proposed by Lennard-Jones and Devonshire
\cite{LJD} and a description of the method can be found in refs.
\cite{HILL,MAGEE02}.  Within the framework of the 3d model, particles
are considered to be localized in singly occupied spherical ``cells'' of
volume $v=V/N$ and radius $s$, centered on the sites of a fully
occupied lattice of some prescribed symmetry. A particle in its cell is
considered to interact with its $c$ nearest neighbors ``smeared''
around the surface of a further sphere of radius $a$ concentric with
the cell. The volume of this ``interaction sphere'' is related to the
cell volume by $a^3=\gamma v\:$ where $\gamma$ is a lattice-dependent
constant, chosen so that for a primitive unit cell of volume $v$ the
lattice parameter will be the radius of the interaction shell .

The Gibbs free energy per particle of the model is given by 

\begin{equation}
\label{cellthA}
g(v)=-k_BT\ln v_{f}\sigma _{c}+\frac{E_0}{2}+Pv\:.
\end{equation}
Here $E_0$ is the ``ground state energy'' - the energy per particle  if
all occupied their lattice sites; $\sigma_{c}$ is a constant ``communal
entropy'' term \cite{FOOTNOTE}, which attempts to account for the
entropy lost due to the localization of particles within cells; and
$v_{f}$ is the ``free volume'',

\begin{equation}
\label{vf}
v_{f}=\int _ve^{-\left(E({\bf r})-E_0\right)/k_BT}d{\bf r}\:,
\end{equation}
with $E({\bf r})$ the ``cell potential'' the interaction
energy of a particle at a position ${\bf r} $ within its cell.
If $c$ neighboring particles are assumed to be smeared
over the interaction shell and the interparticle potential is given by
$u(R)$, the total energy of the particle $E(r)$ is given by

\begin{equation}
\label{E(r)}
E(r)=c\:\frac{\int _{shell}u(R)dA}{4\pi a^2}\:,
\end{equation}
One normally performs this integration numerically with further
numerical integration to calculate the free volume according to
eq.~\ref{vf} and Gibbs free energy for a given choice of number density $\rho=v^{-1}$. 

\begin{figure}[h]
\setlength{\epsfxsize}{11.0cm}
\centerline{\epsffile{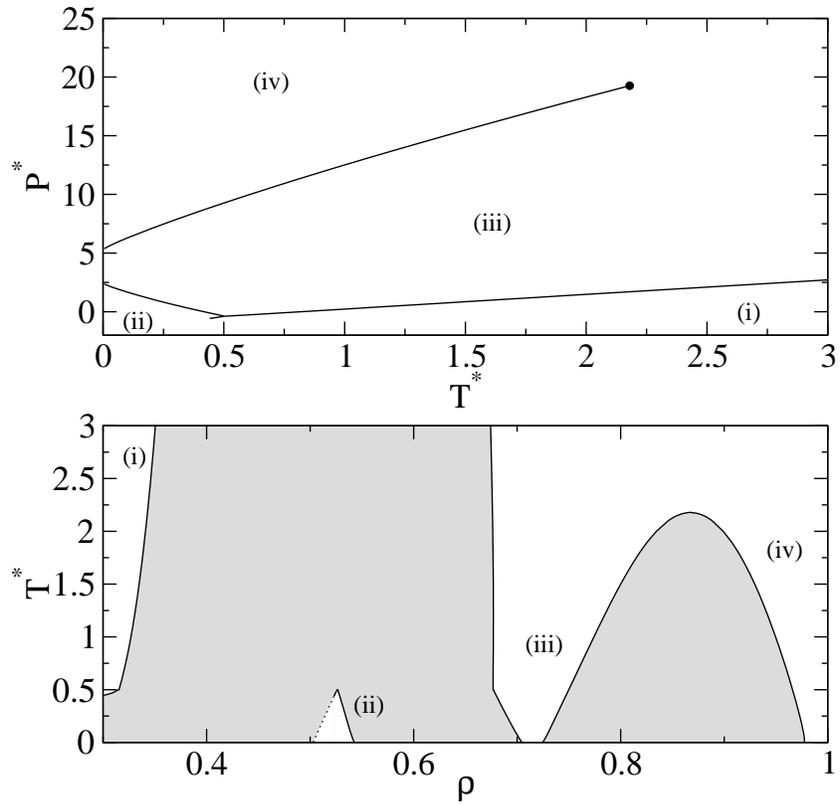}} 
\vspace*{1cm}
\caption{Phase diagram for the 2d should model calculated via the cell theory described in the
text. {\bf (a)} The $P^\star-T^\star$ projection. {\bf (b)} The $T^\star-\rho$ projection.}
\label{fig:tradcell}
\end{figure}

We have obtained the 2d cell theory phase diagram of the potential of
equation~\ref{eq:slpot} in order to compare with the results of
\cite{SL98}. Our study employed a triangular lattice of cells 
 with parameters $c=6$ and $\gamma=2/\sqrt{3}$. The results are shown
in fig.~\ref{fig:tradcell}(a). The phase diagram exhibits four separate
phases, which we have labelled (i)-(iv), delineated by first order
coexistence lines (solid curves). In the case  of phases (i) and (ii),
we were unable to follow the coexistence line right down to zero
temperature, because phase (i) has a density lower than can be
represented by the precision of our calculations. We have therefore
simply extrapolated the measured portion using a linear fit (shown
dotted).

\begin{figure}[h]
\setlength{\epsfxsize}{14.0cm}
\centerline{\epsffile{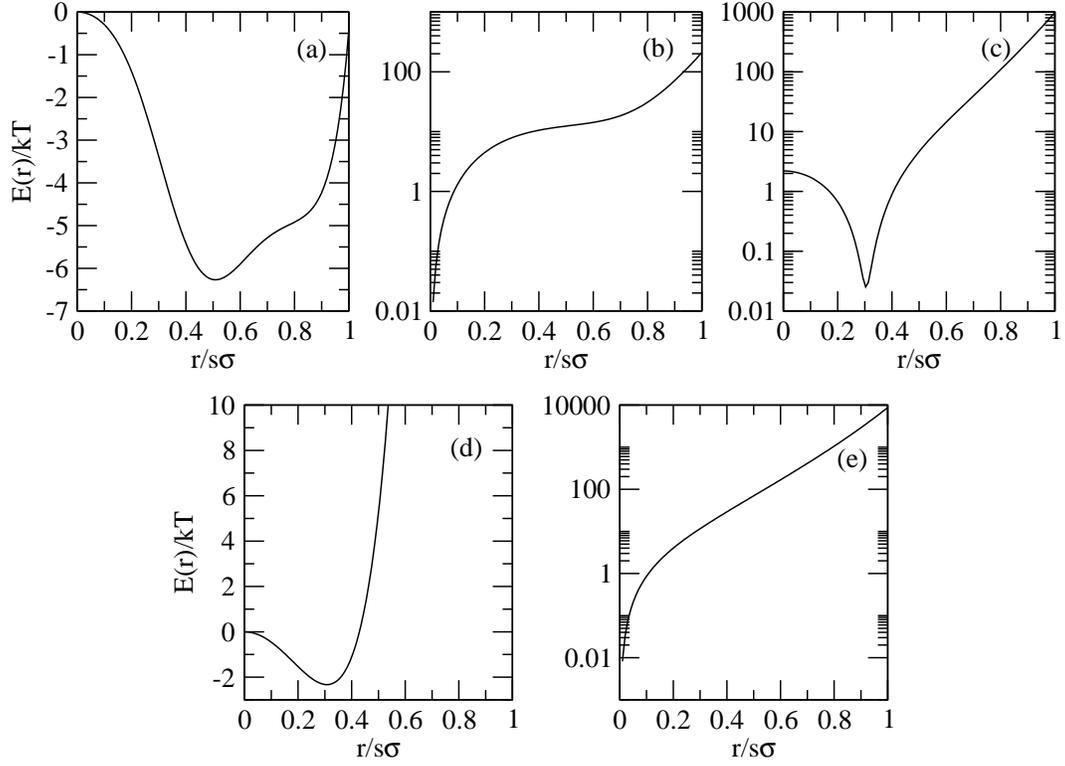}} 
\vspace*{1cm}
\caption{The cell potential, defined in the text, corresponding to the
phases shown in \protect fig.~\ref{fig:tradcell}. {\bf (a)-(c)} correspond to
phases (i)--(iii) respectively, calculated at the triple point. Parts {\bf (d),(e)}
correspond to phases (iii) and (iv) respectively, calculated at coexistence
for $T^\star=0.5$}
\label{fig:cellpotns}
\end{figure}

While no theory based on a prescribed underlying lattice can provide
an unambiguous representation of fluid phases or crystalline solids of
arbitrary structure, a plausible identification of the phases can
nevertheless be made by examining their associated cell potentials
$E(r)$. For phases (i),(ii) and (iii), these have been calculated at
their triple point, permitting a direct comparison under identical
conditions of $P^\star$ and $T^\star$. The results are shown in
fig.~\ref{fig:cellpotns}(a)-(c).  The cell potential at coexistence
between phases (iii) and (iv) are compared in
fig.~\ref{fig:cellpotns}(d)-(e). Inspection of
fig.~\ref{fig:cellpotns}(e) shows that the cell potential for phase
(iv) displays a strong minimum at the cell center $r=0$ (i.e. at the
lattice site). allowing us to identify this phase  as solidlike. By
contrast, for phase (i), the cell potential
(fig.~\ref{fig:cellpotns}(a)) has a non central minimum and, as such,
would be unstable as a lattice phase. Since the density at which the
minimum occurs is low, we tentatively assign this as gaslike. Similar
arguments suggest that phase (iii) is a fluid
(fig.~\ref{fig:cellpotns}(c)) and, because it is denser than phase (i),
and separated from it by a first order phase transition, liquidlike in
character. Additionally, at high temperature (off the scale of our
figure), we find a critical point terminating the first order line of
coexistence between phases (i) and (ii). Finally, the cell potential
for phase (ii) shown in (fig.~\ref{fig:cellpotns}(b)) exhibits a deep
minimum at the lattice site, suggesting it is solidlike.

Given these assignments, a resemblance is evident between
fig.~\ref{fig:tradcell}(a) and that found by simulation
(fig~\ref{fig:TmuPD}). Most strikingly, the freezing transition between
phases (iii) and (ii) displays a negative gradient as found in the
simulations. Moreover, the phase diagram seems to exhibit a second
critical point, terminating the line of coexistence between phases
(iii) and (iv). However, given the previous identification of phase
(iii) as liquidlike and phase (iv) as solidlike, such a second critical
point has no physical counterpart. Furthermore, it should be stressed
that its appearance is {\em not} a unique feature of the CS potential,
as we have recently shown \cite{MAGEE02} in cell theory studies of the
12-6 Lennard-Jones (LJ) potential. Here too,  two critical points we
found; an artifact critical point terminating the liquid-solid
transition and another, a liquid-gas critical point. The latter had not
hitherto been reported in the literature, whilst the artifact critical
point had previously been mistaken for the liquid-gas critical point. 
The appearance of artifact critical points appears to be symptomatic of
the fact that, owing to its lattice-based character, cell theory cannot
properly represent the inherently disordered nature of liquid phases.

\begin{figure}[h]
\setlength{\epsfxsize}{9.0cm}
\centerline{\epsffile{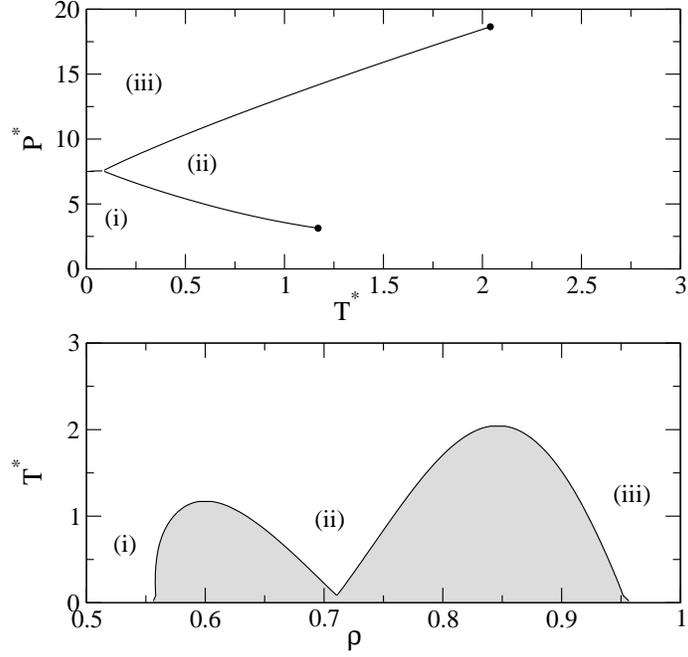}} 
\vspace*{5mm}

\caption{Cell theory phase diagram for the 2d shoulder model,
neglecting the ground state term $E_0$ in the Gibbs free energy (see
text). {\bf (a)} The $P^\star-T^\star$ projection. {\bf (b)} The
$T^\star-\rho$ projection. Phases (i) and (iii) are solidlike, while
phase (ii) is liquidlike.}

\label{fig:nogs}
\end{figure}

We have found no evidence for a fluid-fluid phase transition additional
to the liquid-gas transition in the core softened potential. The
artifact critical point we do find is far removed in the phase diagram
from the second critical point reported in ref~\cite{SL98}. However,
the formulation of cell theory reported there, appears to differ from
the traditional implementation in that it neglects ground state terms
in the potential \cite{YOUNG73}. In view of this, we have also obtained
the phase diagram for this version of the theory. The results
(fig.~\ref{fig:nogs}) display qualitative differences from those of the
more conventional formulation we have described. In particular there
appears to be no liquid-vapor transition. There are, however, two
critical points, although both appear to be artifacts of the cell theory,
one terminating the low pressure solid-liquid coexistence and the other
terminating the high-pressure solid-liquid line. Neither can be
considered to terminate a liquid-liquid transition and neither is
located in the general vicinity of that reported in ref~\cite{SL98}.

\subsection{Solid-solid transitions}

We have investigated the solid phases of the 2d shoulder model as a
function of temperature and pressure using a combination of analytical
and simulation methods. Our motivation for doing so was to check for
the existence of an isostructural solid-solid critical point, such as
that seen in other CS models \cite{KINCAID76,BOLHUIS97}. The large
fluctuations associated with such a critical point (and with the
associated hexatic phases known to occur in 2d systems
\cite{CPhexatic}), might complicate the interpretation of liquid phase
thermodynamic anomalies.

Our investigation of the solid phases begins by performing analytical
calculation based on the harmonic approximation (HA). These supply
results which, whilst exact in the low-temperature limit, lose accuracy
with increasing temperature. They are thus used as a starting point for
direct simulations of lattice-lattice phase coexistence using the
Lattice Switch Monte Carlo (LSMC) method \cite{LSMC1,LSMC2,softLSMC}.
Whilst powerful, we found that for this particular model LSMC becomes
inefficient as melting is approached owing to extended sampling times
caused by a significant defect concentration. Thus we turn at high
temperature to Gibbs-Duhem integration which provides a faster, though
less accurate approach to tracing coexistence curves.

Both the HA and LSMC approaches are designed for studying solid phases.
However they share a common problem, namely that rather than predicting
which of the set of possible crystal lattice occurs, a chosen subset of
lattices must be proposed and checked against one other for relative
stability. In this work, we only consider the possibility of square and
triangular (hexagonal) lattices, these being the ones observed in
the original work \cite{SL98}. It is, nevertheless, possible that other
stable lattices may exist (see ref.~\cite{JAGLA98} for an example).

\subsubsection*{Harmonic approximation studies}

\begin{figure}[h]
\setlength{\epsfxsize}{14.0cm}
\centerline{\epsffile{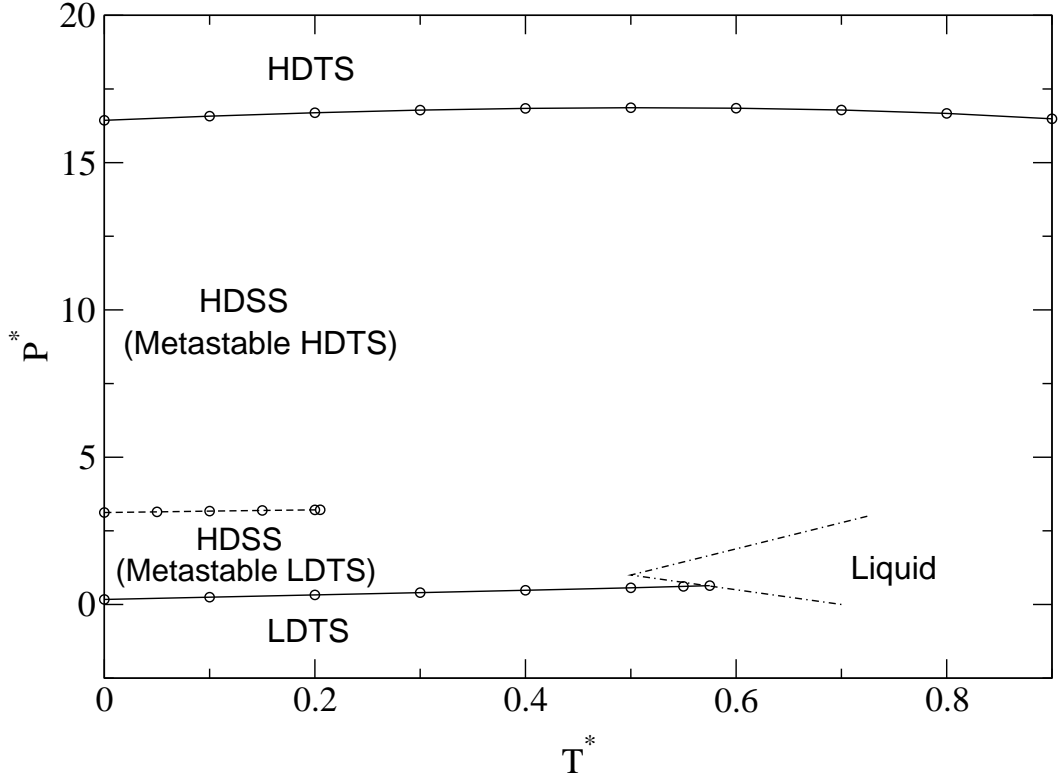}} 
\vspace*{5mm}

\caption{$P^\star-T^\star$ projection of the 2d shoulder model phase
diagram calculated within the harmonic approximation. Circles indicate
the state points at which calculations were performed, curves guide the
eye. Solid lines indicate transitions between thermodynamically stable
phases. The dashed line represents the locus within the stable square
lattice phases at which the metastable phase changes from being the
LDTS to the HDTS. The dot-dashed line shows the melting curve reported
in reference~\protect\cite{SL98}.}

\label{haphdiag}
\end{figure}

\begin{figure}[h]
\setlength{\epsfxsize}{9.0cm}
\centerline{\epsffile{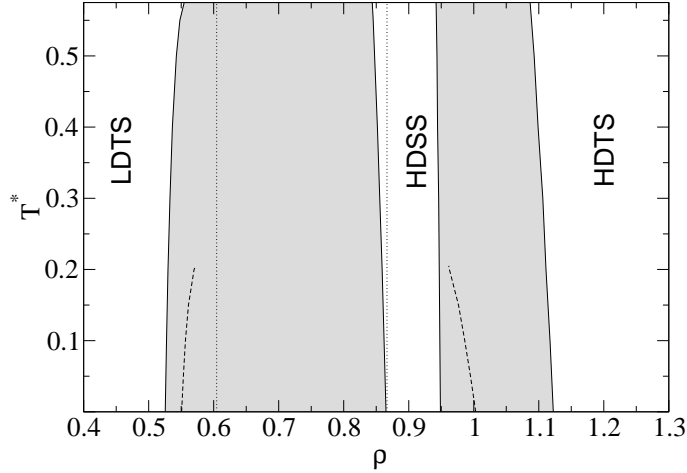}} 
\vspace*{5mm}

\caption{ $T^\star-\rho$ projection of the 2d shoulder model phase
diagram calculated within the harmonic approximation. Solid
lines indicate the equilibrium coexistence densities between phases.
Dashed lines indicate the binodal for the metastable LDTS-HDTS
transition. Dotted lines correspond to the spinodals of the HDTS and
LDTS phases}
 
\label{haVTphdiag}
\end{figure}

Harmonic approximation (HA) calculations \cite{A&M} have been performed
for a $16\times 16$ system of particles on both square and triangular
lattices. Coexistence lines were located by using a Newton-Raphson
root-finding algorithm \cite{NRC} to solve for conditions of equal
pressure and Gibbs free energy. The results are presented in figs.
\ref{haphdiag} and \ref{haVTphdiag} and show three transition lines. At
very low pressure, an open triangular lattice phase (LDTS) is stable,
which transforms to a dense square lattice phase (HDSS) on increasing
pressure, in line with the results of ref.~\cite{SL98}. At very high
pressure, this dense square lattice undergoes a transition to a new
dense triangular lattice phase (HDTS). The third transition line occurs
at intermediate pressure, and is isostructural, separating LDTS and
HDTS phases, both metastable with respect to the square lattice phase.

The existence of a stable square lattice phase is unusual. In the case
of the shoulder potential, however, the lattice has nearest neighbors
sitting in the minimum of the Lennard-Jones part of the interaction
potentials, and is stabilized through second-nearest neighbors sitting
in the deep Gaussian well part (cf. fig.~\ref{fig:hds_conf}(c) and
fig.~\ref{fig:pots}). We find that if second-nearest neighbor
interactions are ``turned off'', the presence of the square lattice is
completely suppressed.

We have calculated the coexistence lines for reduced temperatures up to
$T^\star=0.9$. The ``hidden'' isostructural line finishes at 
$T^\star=0.205$, whilst the LDTS-HDSS line ends at $T^\star=0.575$. In
both these cases, the transition line ends because the root finding
algorithm cannot identify a coexistence volume for the LDTS phase which
is mechanically stable within the approximation. In the case of the
LDTS-HDSS transition, this probably indicates approach to melting and
the associated breakdown of the approximation. In the case of the
hidden isostructural transition, this could simply be a point beyond
which the LDTS phase loses mechanical stability, or it could indicate
approach to a hidden isostructural critical point (the associated
fluctuations of which would also cause breakdown of the approximation).
If there is a hidden isostructural critical point, with any associated
region of hexatic phase stability \cite{CPhexatic}, we note that it
would be at far too high a pressure to be in any way associated with
the thermodynamic anomalies noted in the liquid phase.

The HDTS-HDSS coexistence line continues across the region we have
checked; its only peculiarity is that it passes through a pressure
maximum at around  $T^\star\approx 0.51$. From the Clausius-Clapeyron
equation,  $\frac{dP}{dT}=\frac{\Delta S}{\Delta V}$, we know that
passing through this point with increasing temperature, the entropy
difference between the HDSS and HDTS phases changes sign. We suggest
that at low temperatures (where particles remain close to their lattice
sites) the greater entropy of the square lattice reflects its larger
volume. With increasing temperature, however, particles in the
triangular lattice will be free to explore an ever greater region of
configuration space,  whilst those in the square lattice will be
constrained by the conditions on the positions of second-nearest
neighbors necessary to maintain mechanical stability. Accordingly the
square lattice will have lower entropy.

\subsubsection*{Lattice Switch Monte Carlo and Gibbs-Duhem integration studies}

At high temperature, or on approaching the melting transition, the HA
breaks down. We have therefore used our HA results as the starting
point for a method which uses direct two-phase simulation of coexisting
solid phases - Lattice Switch Monte Carlo (LSMC)
\cite{LSMC1,LSMC2,softLSMC}. Our implementation of LSMC is similar to
that described in reference \cite{softLSMC}, except that we operate in
the constant-$NPT$ ensemble \cite{FRENKEL}, and as such our order
parameter is the difference in enthalpy, $\Delta H$, between conjugate
pairs of configurations. In addition to alternating between different
sets of lattice vectors, our lattice switch move also alters the aspect
ratio of the box between $ 1:1 $ (for a square lattice) and $
\sqrt{\frac{\sqrt{3}}{2}}:\sqrt{\frac{2}{\sqrt{3}}} $ (for a triangular
lattice)\cite{NOTE1}, and adds a lattice-dependent scaling factor to
move between the different characteristic volumes for each phase
(determined by two short single-phase simulations at the state point).

LSMC simulations were performed for both the LDTS-HDSS and HDSS-HDTS
coexistence lines for a system of $256$ particles. Whilst metastable
triangular phases of the correct volumes for the hidden isostructural
transition were found at low pressure, their lifetime in a Monte Carlo
run was insufficient for simulation of coexistence. For the HDSS-HDTS
transition, simulations were performed at two low-temperature state
points on the phase boundary to verify its existence; coexistence was
found at $T^\star=0.1, P^\star=16.60$ and $T^\star=0.2, P^\star=16.73$.
These results agree well with those from the HA. 

\begin{figure}[h]
\setlength{\epsfxsize}{14.0cm}
\centerline{\epsffile{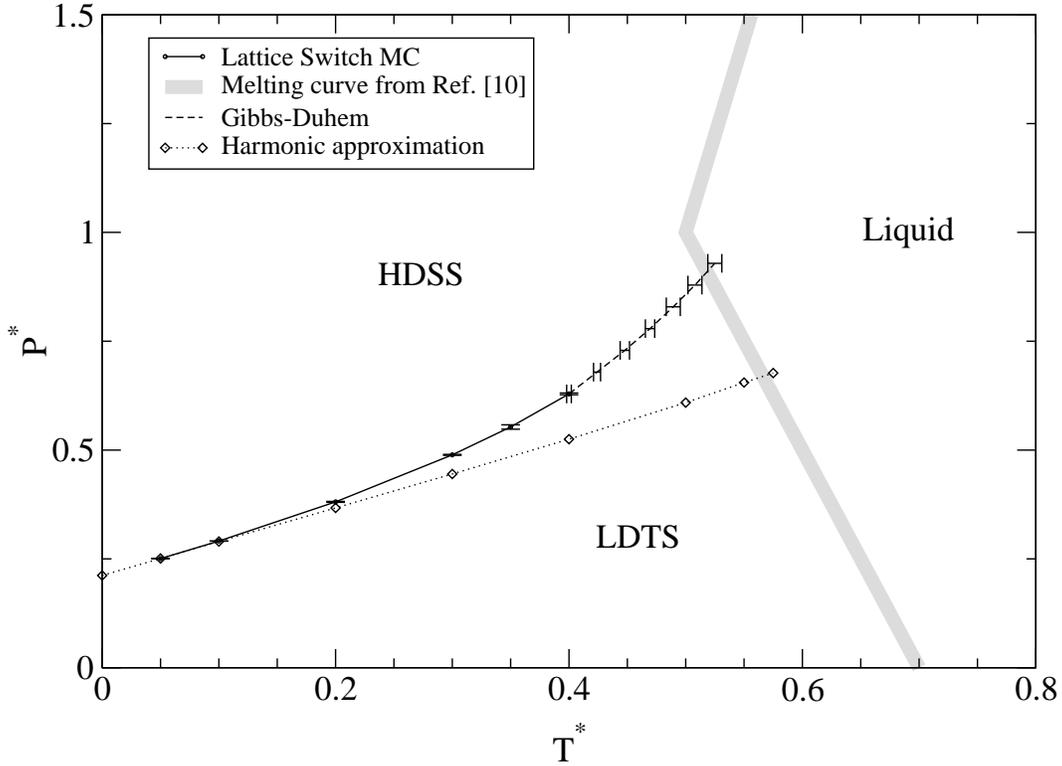}} 
\vspace*{5mm}

\caption{Phase diagram showing the HDSS-LDTS transition line calculated
within the harmonic approximation (dotted line), and from LSMC
simulations (full line) and from Gibbs-Duhem integration (dashed line).
Also shown is the melting curve taken from reference
\protect\cite{SL98} (thick grey line).} \label{fullTP}

\end{figure}

Results from our LSMC simulations for the LDTS-HDTS line are shown in
fig.~\ref{fullTP}. Up to a reduced temperature of $T^\star=0.2 $, these
agree well with the HA results, but above this temperature the
simulation coexistence line lies at higher pressures than the HA
coexistence line. This is to be expected, since the HA is unlikely to
deal well with the very low density of the LDTS phase.

As the temperature was raised above $T^\star=0.2$, the LSMC method was
found to become increasingly less efficient. An essential aspect of the
method is to bias the phase space sampling such that the system
regularly samples configurations in which the particles are very close
to their lattice sites. As the melting temperature was approached, this
became progressively more problematic, due to an increase in the defect
concentration. Indeed the presence of such defects is well known to be
a feature of 2d melting \cite{2DMELT}. Once the reduced temperature $
T^\star=0.4 $ was reached, the problem became so severe that use of
LSMC was no longer feasible. It was therefore decided to continue
tracing the coexistence curve to higher temperatures using Gibbs-Duhem
(G-D) integration \cite{G-D,FRENKEL}. This involves performing separate
single phase simulations at a coexistence state point, from which the
slope of the coexistence curve is estimated via the Clausius-Clapeyron
equation. Integration of this equation allows the locus of the
coexistence line to be tracked. The G-D method assumes nothing about
the coexisting phases apart from that they are (meta)stable over the
timescale of the simulations, and is both efficient and easy to
perform. Our G-D implementation employed a simple two-step trapezoid
predictor-corrector integrator \cite{NRC}; the results are shown as the
dashed line in fig.~\ref{fullTP}. We see that the G-D estimated
coexistence line smoothly extends the LSMC line, up to a temperature $
T^\star=0.525 $; above this temperature, the triangular lattice phase
was observed to melt. This coexistence line terminates at a temperature
which agrees (within error) with the melting curve estimated by
Sadr-Lahijany {\em et al} ~\cite{SL98}, but at a pressure significantly
less than their estimate for the triple point. 

\section{Ramp model}

Jagla \cite{JAGLA01} has recently presented a MC simulation study of a
3d CS fluid described by the potential

\begin{eqnarray} 
\displaystyle
U(r)= & \infty\hspace*{4.5cm} & r < r_0\:, \nonumber \\
U(r)= & \displaystyle\epsilon\frac{(r_1-r)}{(r_1-r_0)}-\gamma\frac{(r_2-r)}{(r_2-r_0)}\hspace{1cm} & r_0 \leq r< r_1\:, \nonumber \\
U(r)= & \displaystyle-\gamma\frac{(r_2-r)}{(r_2-r_0)} \hspace{3cm} & r_1 \leq r< r_2\:, \nonumber \\
U(r)= & 0 \hspace*{4.5cm}  & r\geq r_2\:,
\label{eq:jagpotdef}
\end{eqnarray}
with $r_1=1.72r_0, r_2=3.0r_0, \gamma=0.31\epsilon$. The form of this
potential is depicted in fig.~\ref{fig:pots}.

Using canonical ensemble (constant-NVT) MC simulation, Jagla  obtained
the pressure as a function of system volume for a 3d system of $300$
particles. Van der Waals loops were observed in the measured
$P(V)$ curves, suggestive of the existence of a phase
separation between a low density liquid (LDL) and a high density liquid
(HDL). The critical point of this transition (measured as the
temperature above which the loops disappear) was reported to occur at
the low temperature of $T^\star\approx 0.08$. Additionally an anomalous
increase in the density of the low density liquid was observed with
increasing temperature. 

A loop in a $P(V)$ curve below the critical temperature, instead of a 
flat region, is a finite-size artifact of the constant-$NVT$ ensemble
and as such cannot be regarded as an unambiguous indicator of a phase
transition. In view of this, we have attempted to corroborate Jagla's
findings for the 3d ramp model in greater detail, using simulation
methods designed for studies of fluid phase coexistence. Initially we
sought to perform GCE simulations, close to the reported location of
the LDL-HDL critical point. Unfortunately, these proved extremely
inefficient due to a very low acceptance rate for particle transfers at
these parameters. The results we report here instead derive from
simulations within the isothermal-isobaric (constant-$NPT$) ensemble
\cite{FRENKEL} which, in the low temperature regime, proved
considerably more efficient than the GCE. Notwithstanding its greater
efficiency, use of the constant-$NPT$ ensemble did not permit the study
of very large system sizes. This problem is traceable to the method's
general inefficiency in the context of hard-core fluids,
and stems from the necessity of rejecting all proposed volume
contractions that result in a hard core overlap.  Consequently, we were
able to study only three system sizes, comprising $N=300$, $N=500$ and
$N=800$ particles respectively.

\begin{figure}[h]
\setlength{\epsfxsize}{10.0cm}
\centerline{\epsffile{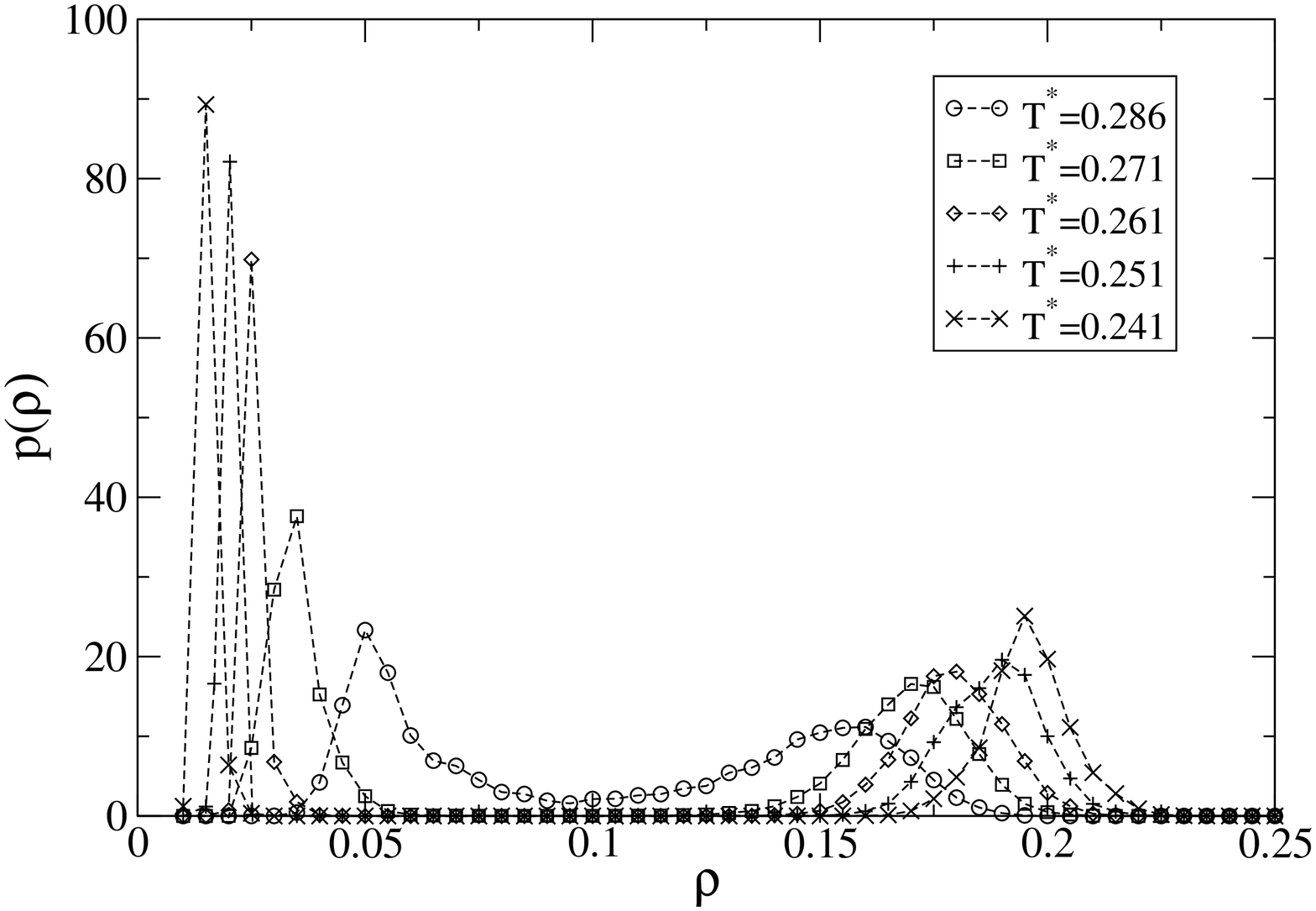}} 
\setlength{\epsfxsize}{10.0cm}
\centerline{\epsffile{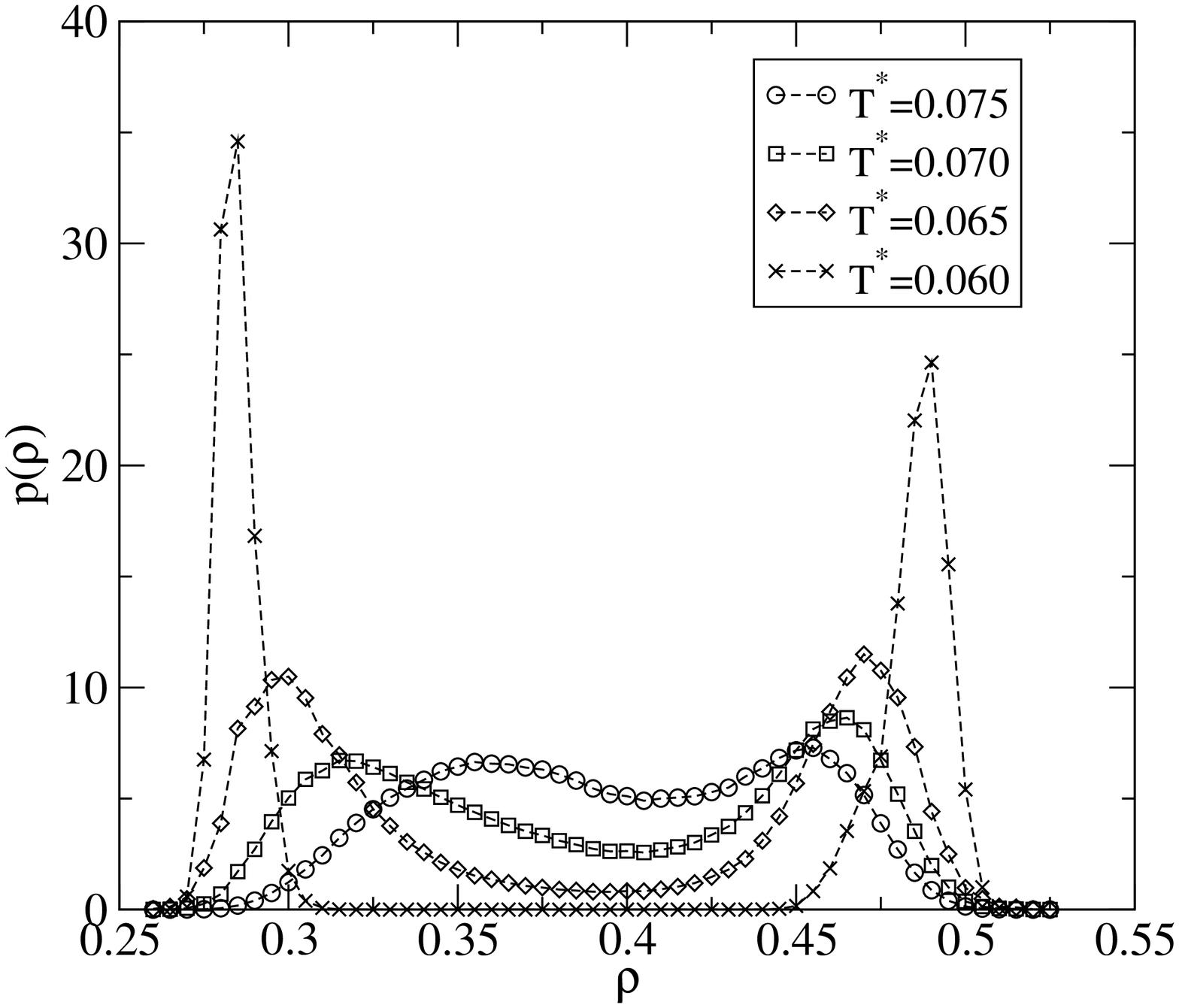}} 
\vspace*{5mm}
\caption{The measured coexistence forms of the number density
distribution obtained in the manner described in the text, for $N=300$. {\bf
(a)} the LDL-gas coexistence boundary. {\bf (b)} The LDL-HDL
boundary. Lines merely serve as guides to the eye. }
\label{fig:jagdists}
\end{figure}

\begin{figure}[h]
\setlength{\epsfxsize}{10.0cm}
\centerline{\epsffile{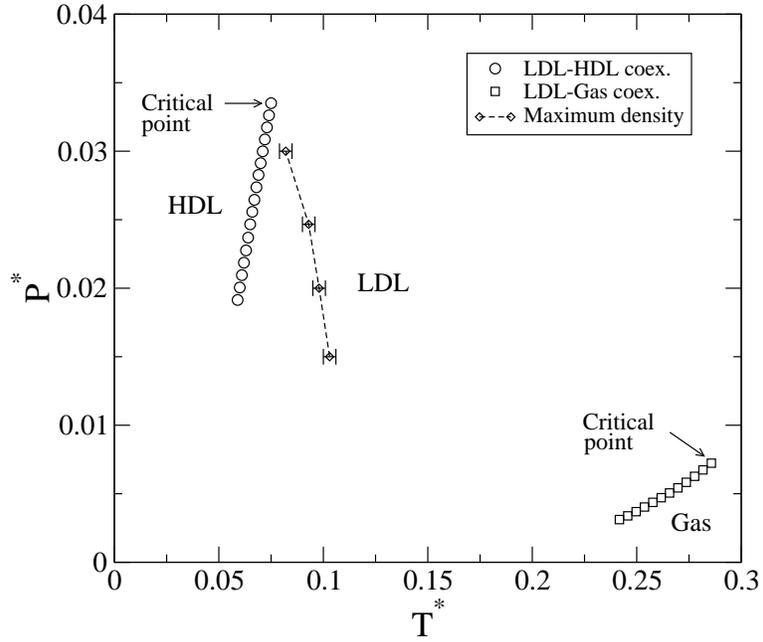}} 
\vspace*{5mm}

\caption{The liquid-gas and LDL-HDL coexistence lines in the
$P^\star-T^\star$ plane obtained for the $N=300$ system size.
Statistical uncertainties are smaller than the symbol sizes. Also shown
($\diamond$) is the locus of the line of maximum density in the LDL
phase.}

\label{fig:jagpd}
\end{figure}

Our studies revealed two fluid-fluid phase boundaries. One, a
liquid-gas line occurs at high temperature and low pressure; the other
separates an LDL phase from a HDL phase, as previously reported by
Jagla \cite{JAGLA01}.  To track these boundaries, we utilized
multicanonical sampling and histogram reweighting techniques in the
manner described in ref.~\cite{WILDING95} to yield the coexistence
forms of the density distribution $p(\rho)$ (cf.
sec.~\ref{sec:sublimation}). The tracking procedure was initialized
near the critical point and followed the phase boundary down in
temperature until the simulations became too slow to continue.  A
selection of coexistence density distributions from each phase boundary
is shown in figure~\ref{fig:jagdists}. The associated phase diagram in
the $P^\star-T^\star$ plane appears in fig.~\ref{fig:jagpd}. Matching
to the known universal form of the order parameter distribution
\cite{BRUCE92,WILDING95} allows us to estimate the critical parameters.
The LDL-gas critical point lies at $T_C^\star=0.2857(3),
P_c^\star\equiv P\sigma^2/\epsilon=0.00723(1)$, and has an unusually
low critical density of $\rho_c=0.10(1)$. The critical point of the
LDL-HDL boundary lies at $T_c^\star=0.076(2),
P_c^\star=0.0341(5),\rho_c=0.378(3)$. We note that to within the
precision of our measurements, the LDL-HDL phase boundary is linear,
while the LDL-gas phase boundary exhibits a positive curvature. It is
also apparent that the two phase boundaries are very well separated in
the phase diagram.  Since we were unable to probe the region of very
low temperature, we cannot say whether or not there exists a triple
point between the gas, low density liquid and the high density liquid.
It seems more likely that both phase boundaries terminate in solid
phases. 

\begin{figure}[h]
\setlength{\epsfxsize}{10.0cm}
\centerline{\epsffile{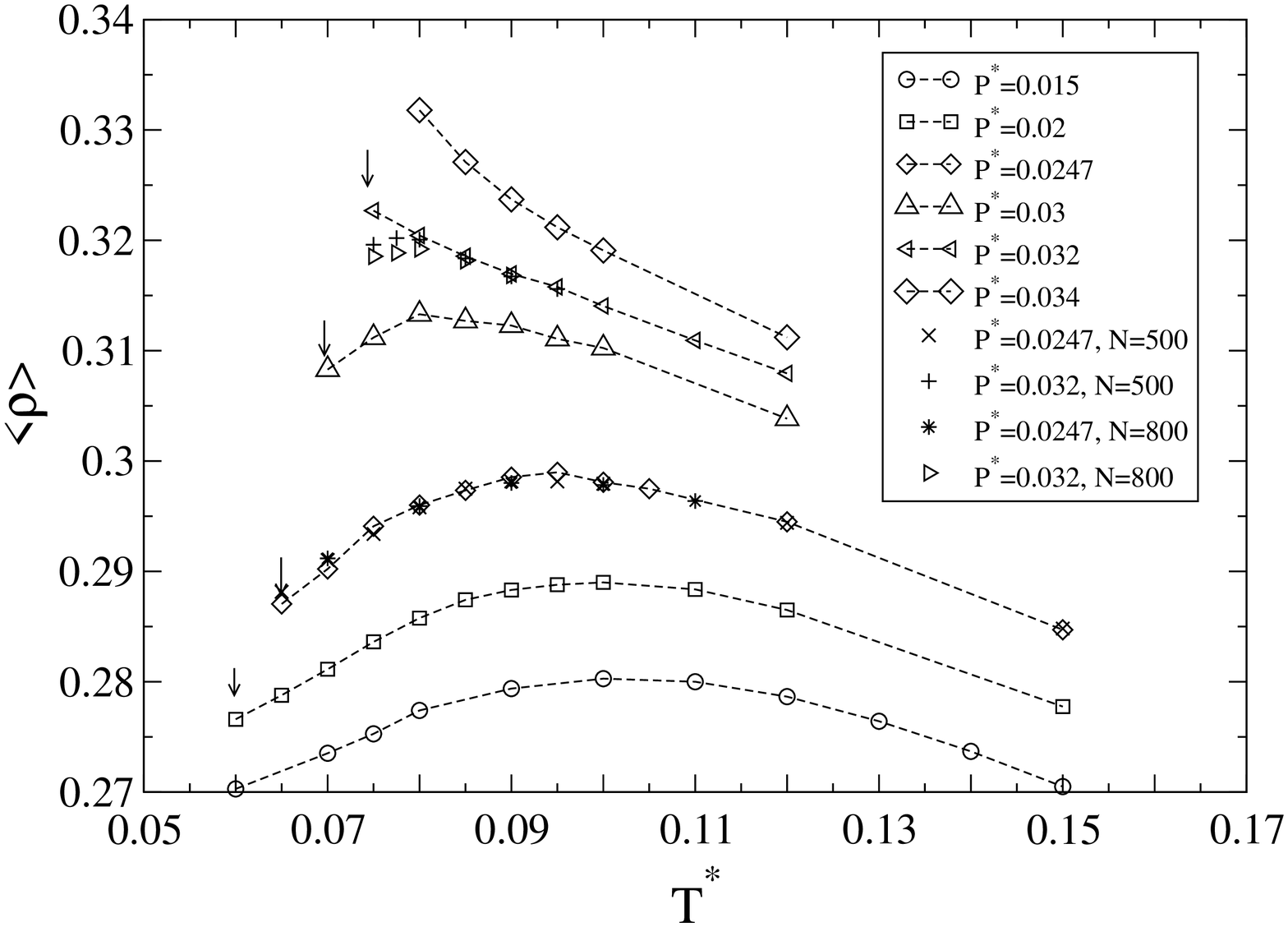}} 
\setlength{\epsfxsize}{10.0cm}
\centerline{\epsffile{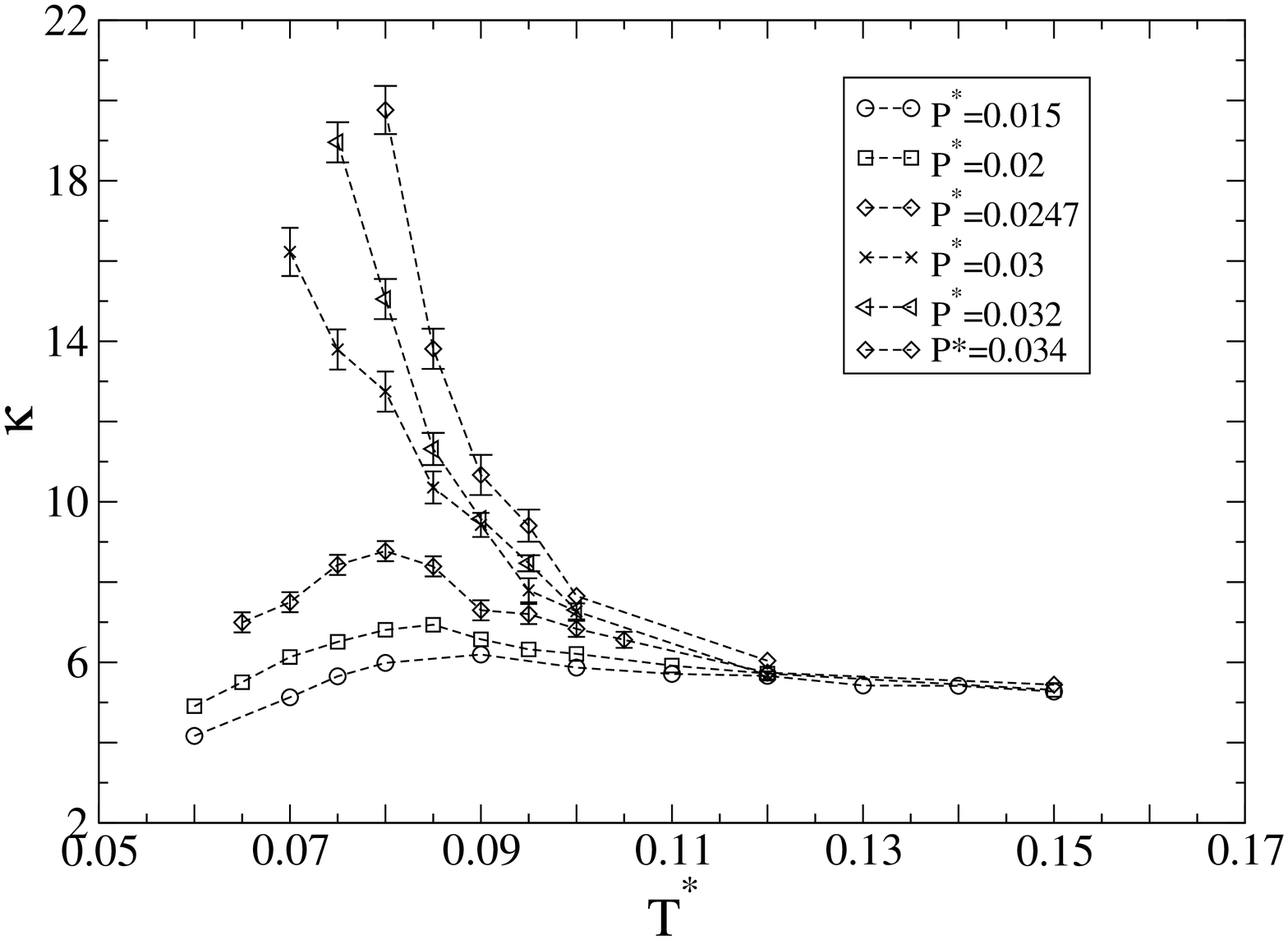}}
\vspace*{5mm}

\caption{{\bf (a)} The temperature dependence of the average number
density along selected isobars for $N=300$. Data is also shown for
$N=500$ and $N=800$ at selected pressures.  Lines are merely guides to
the eye. Arrows indicate the LDL-HDL coexistence temperature for each
isobar.  {\bf (b)} The corresponding estimates of the compressibility
for $N=300$. Unless otherwise shown, the magnitude of statistical
errors does not exceed the symbol sizes.}

\label{fig:jaganom}
\end{figure}

Jagla \cite{JAGLA01} observed thermodynamic anomalies in his $P-V$
curves of the LDL phase, but did not trace their path through the phase
diagram. We have sought to do so by measuring the number density and
compressibility along isobars, starting at the LDL-HDL transition and
increasing in temperature. The results can be seen in
fig.~\ref{fig:jaganom}(a) and (b) respectively. One observes that for
$P^\star\ll P_c$  there is both a density and a compressibility maximum
at temperatures well in excess of the LDL-HDL coexistence values. Well
away from the critical point (at $P^\star=0.72P^\star_c$), comparison
of the data for the three system sizes indicates no significant
finite-size effects in the form of the density maximum \cite{FSE}.

On increasing the pressure towards its critical value, the height of
the compressibility peak grows. Additionally, the temperature at which
the maxima occur shifts closer to the coexistence curve. In this regime,
we do see finite-size differences in the form of the density maximum.
In particular, the density maximum which is not visible at
$P^\star=0.032$ for $N=300$, reappears very close to the coexistence
curve for $N=500$ and $N=800$ (fig.~\ref{fig:jaganom}(a)). The line of
density maxima as a function of pressure for $N=300$, is included on
the phase diagram of fig.~\ref{fig:jagpd}. The line intersects the
coexistence curve slightly below the second critical point. In view of
the observed finite-size dependence of the density maximum, it seem
likely that in the thermodynamic limit, the maximum will intersect the
liquid-liquid line even closer to the critical point, and might even
converge on the second critical point itself. It should be stressed
however, that this can only be confirmed via a full finite-size scaling
analysis of the critical region, the computational requirements of
which exceeded the available resources.

As regards the temperature behaviour of the compressibility, obtaining
good statistics for this quantity is harder, because it is measured via
the second moment of the density distribution. However, in contrast to
the density, its maximum clearly vanishes well before the critical
pressure is reached (fig.~\ref{fig:jaganom}(b)). A likely explanation
for this difference is to be found in the fact that the compressibility
exhibits a critical point divergence. This presumably swamps the
anomalous compressibility maximum well before the critical point is
reached.

\section{Summary, discussion and conclusions} 
\label{sec:concs}

In this paper we have investigated the phase behaviour and liquid state
anomalies of two distinct CS models. Below, we summarize and discuss
our results for each model in turn.

For the 2d shoulder potential, we obtained the liquid-gas coexistence
curve and studied how it evolves into the LDTS sublimation line at the
gas-liquid-LDTS triple point. In the solid region of the phase diagram
we employed Lattice Switch MC techniques and Gibbs-Duhem integration to
map the phase boundary between the LDTS and the HDSS from very low
temperatures up to the melting point. Analytical calculations within
the harmonic approximation suggested the existence of a reentrant
triangular solid phase at very high pressures, the HDTS.  The existence
of this phase was confirmed using Lattice Switch MC. Evidence was also
found for a metastable LDTS-HDTS and associated critical point lying at
low temperatures and high pressures, well below LDTS melting temperature.

Considerable effort was devoted to probing the behaviour of the
liquid state anomalies in the number density and compressibility in the
liquid phase. Maxima in the density and compressibility were observed
along two widely separated isobars. An inspection of configurations
either side of the compressibility maximum indicated that its presence
is associated with the freezing of the liquid to the LDTS. The double
peaked nature of the density distribution function at the temperature
of the compressibility maximum confirmed the existence of two distinct
favored regions of density. 

Measurements of the temperature dependence of the compressibility and
density maxima for a wide range of system sizes revealed pronounced
finite-size effects. For small to medium system sizes, these took the
form of systematic changes in the peak heights and peak temperatures.
Interestingly, however, the differences between the results for the
largest systems were much less pronounced than for the smallest
systems. Finite-size difference between the smallest and largest
systems, of the magnitude observed, are strongly indicative of a
large correlation length $\xi$. The apparent convergence of the
results for the compressibility peak suggests that this correlation
length is {\em large but finite} at the transition. In such
circumstances one would expect that in the regime $L < \xi$,
finite-size effects are great, while for system sizes $L\gtrsim \xi$,
they begin to die away. In support of this, we remark that it is widely
believed that the 2d freezing transition has a pseudo-continuous
character, although the precise nature of the transition remains the
subject of some debate \cite{2DMELT,SOMER98}.

\begin{figure}[h]
\setlength{\epsfxsize}{9.0cm}
\centerline{\epsffile{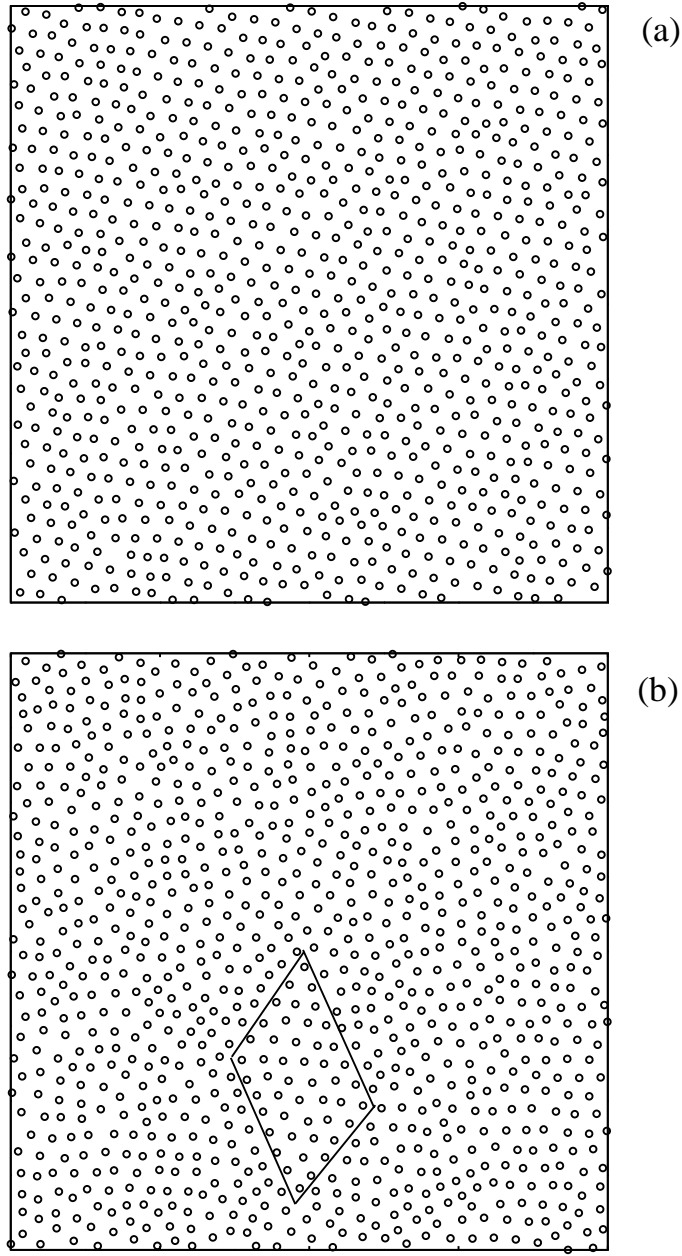}} 
\vspace*{5mm}
\caption{Typical snapshot configurations taken from the $L=35\sigma$
system at $\mu^\star=-3.0$. (a) $T^\star=0.55$, the system is in the LDTS phase (b)
$T^\star=0.6$, the system is in a liquid-like phase, but displays clusters of
crystalline ordering, one of which is ringed.}
\label{fig:freezeconfs}
\end{figure}

\begin{figure}
\setlength{\epsfxsize}{9.0cm}
\centerline{\epsffile{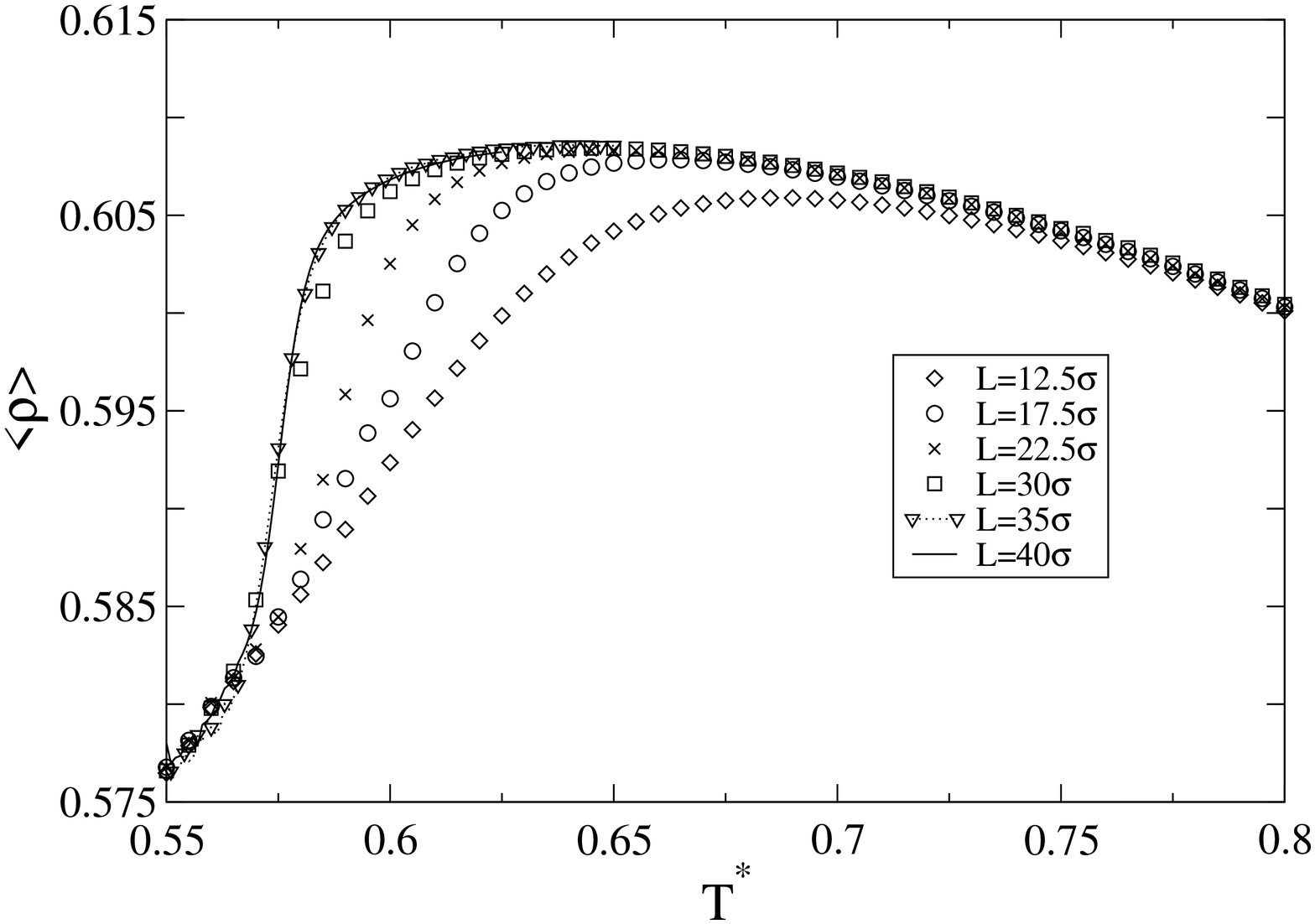}} 
\setlength{\epsfxsize}{9.0cm}
\vspace*{1cm}
\centerline{\epsffile{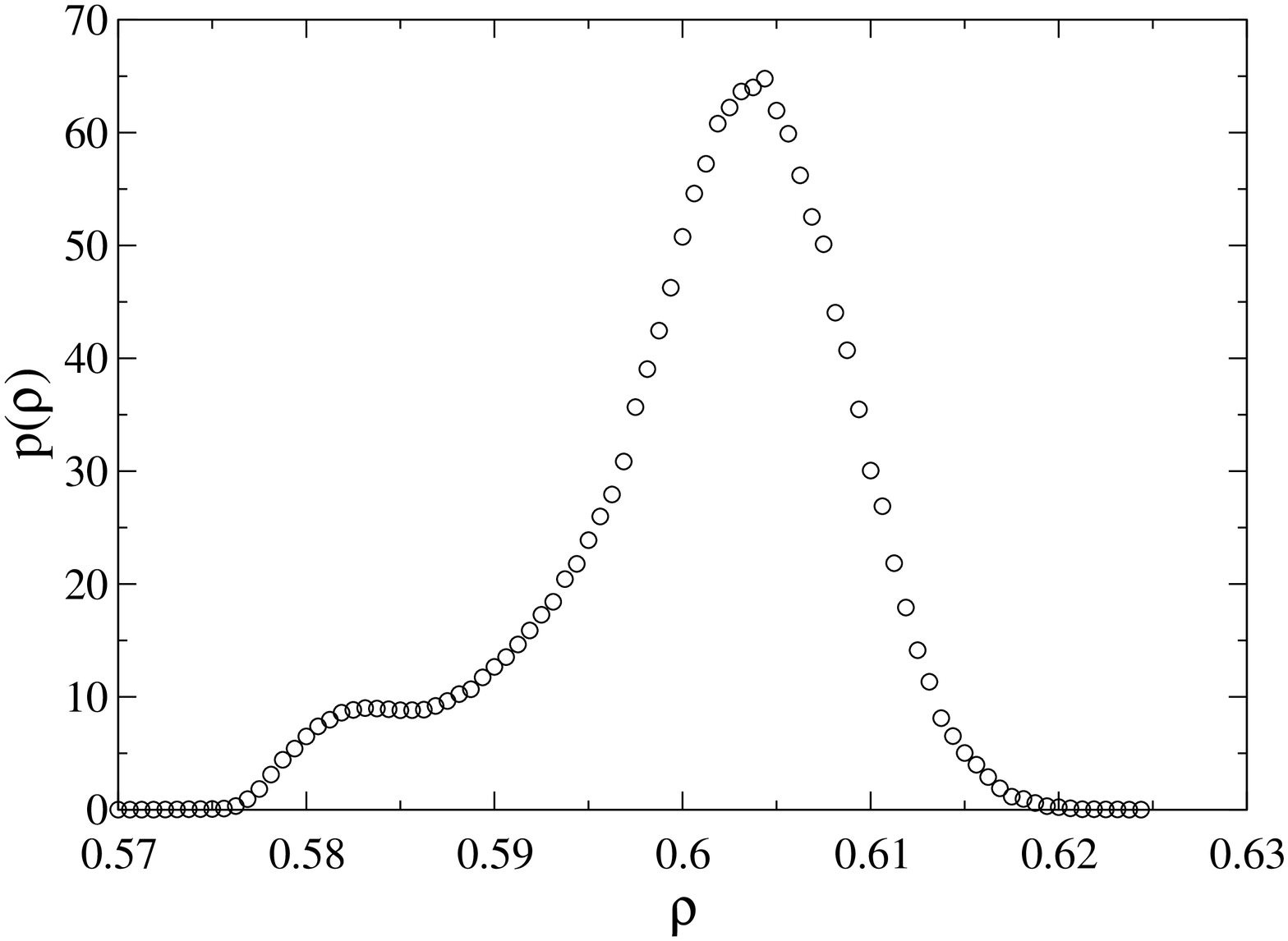}} 
\vspace*{5mm}

\caption{{\bf (a)} The measured temperature-dependence of the average
density for $\mu^\star=-3.0$, for a range of system sizes. {\bf (b)}
Number density distribution, $p(\rho)$, for the $L=40\sigma$ system size at
$\mu^\star=-3.0,\: T^\star=0.58$, corresponding to the point where the
density curve starts to fall in ({\bf a)}.}

\label{fig:den_mu3.0}
\end{figure}

Our results for the density distribution $p(\rho)$ suggest that the
density maximum, (whilst clearly occurring at temperatures in excess of
the freezing temperature), is associated with the first appearance of
clusters of LDTS crystalline structure within the liquid phase \cite{TDL}. This is
evidenced by the subsidiary peak occurring at the LDTS density, (cf. 
fig.~\ref{fig:den_mu3.0}(b)) and by the liquid phase snapshots just
above the melting temperature, fig.~\ref{fig:freezeconfs}(b), at the 
point where the density starts to fall rapidly. Since these solidlike
clusters possess a lower local density than that of the coexisting
liquid, their presence reduces the average density. As the temperature
is lowered towards the freezing transition, the size and persistence of
the solid-like clusters increases, causing the average density to fall
ever more rapidly, (cf. fig.~\ref{fig:den_mu3.0}(b)). The shift of the
peak in the average density to lower temperatures with increasing $L$
is a natural consequence of the fact that for $L\lesssim \xi$, the
apparent freezing point will occur at higher temperatures than in the
thermodynamic limit. 

We have also addressed reports of a liquid-liquid critical point within
cell theory for the 2d shoulder model. Our own cell model study
(sec.~\ref{sec:cell}), does reveal a second critical point, but at
parameters far removed from those quoted in ref. \cite{SL98}. We
stress, however, that the appearance of a second critical point within
cell theory is not a feature peculiar to CS potentials. Indeed in a
recent reappraisal of cell theory for the $12-6$ Lennard-Jones fluid
\cite{MAGEE02}, we have shown that this model also exhibits a second
critical point. Moreover, the second critical point occurring in both
the Lennard-Jones and CS models appears to be an artifact of the
lattice-based nature of cell theory, terminating as it does a phase
boundary between a liquidlike and a solidlike phase. In view of this it
seems unlikely that cell theory will ever represent a useful approach
for investigating the liquid phase behaviour of CS systems. 

Taken together, our results lead us to conclude that the thermodynamic
anomalies of the 2d shoulder model are not caused by strong
liquid-state fluctuations associated with the proximity of a
liquid-liquid critical point (metastable or otherwise). Instead we
attribute them to the strong pseudo-critical fluctuations associated
with the 2d freezing transition and the fact that the density of the
LDTS is lower than that of the liquid. Whilst we do not discount the
possibility of a metastable liquid-liquid critical point somewhere
within the LDTS phase, it is difficult to see how it could (in the
interpretation of ref. \cite{SL98}) give rise to a compressibility
maximum of the scale we observe on two such widely separated isobars.
We further remark that our interpretation of the origin of the
anomalies is consistent with the recently reported failure to observe
their presence in 3d versions of shoulder models
\cite{BULDYREV02,FRANZESE_unpub}. Here the freezing transition is
expected to be sharply first order in character, with no large
fluctuating clusters of the solid phase within the liquid. As such
there will be little or no sign of the approach to freezing.

Turning now to our results for Jagla's ramp potential in 3d, we find a
liquid-gas coexistence curve at high temperature and low pressure, and
confirm the existence of a stable transition between a high density
liquid and a low density liquid at lower temperature and higher
pressure. For both transitions we have mapped a portion of the
coexistence curve and determined the critical parameters. Within the
LDL phase, and below the temperature of the second critical point, we
find maxima in the density and compressibility as a function of
temperature. In contrast to the 2d shoulder model, these anomalies are
authentic, i.e. they are not associated with the formation of an
incipient phase. The locus of density maxima appears to converge on
the second critical point. 

Finally, with regard to the general issues raised by our findings, the
presence of a stable second critical point in the ramp model begs the
question as to what features of this potential are responsible for its
existence, when none is observed in other single component fluids such
as the Lennard-Jones fluid. Although we have not yet studied this
matter in detail, it seems likely that the potential minimum must be
located at a sufficiently large radius relative to the hard core,
otherwise the LDL-HDL transition is preempted by the freezing
transition. This mirrors the known requirements for the existence of a
liquid-gas transition, which is stable with respect to the crystalline
phases only for an attractive potential of sufficiently large range
\cite{FRENKEL00}. It would be of considerable interest to examine the
precise role of the interaction range on the stability of the LDL-HDL
critical point, as well as other factors such as the steepness of the
soft core and the depth of the minimum. Such a study could benefit the
general understanding of the relationship between waterlike anomalies
in real systems and second critical points, whether metastable or
otherwise \cite{NOTE2}. We hope to report on this matter in future
work. 

\acknowledgements

The authors thank A.D. Bruce for numerous helpful conversations and
suggestions. NBW is grateful to G. Stell for introducing him to the
topic of core-softened potentials. JEM acknowledges an EPSRC
studentship.

\newpage

\newpage

\newpage

\newpage

\newpage

\newpage

\newpage

\begin{thebibliography}{99}
\bibitem{FOOTNOTE} Reference \protect\cite{LJD} assigns $\sigma_c=e$,
although the exact value chosen for $\sigma _{c}$ has no impact upon
phase coexistence so long as it is independent of volume.
\bibitem{NOTE1} This aspect ratio change, which leaves the volume
unaltered, is necessary since a triangular lattice is not commensurate
with a square box.
\bibitem{FSE} This suggests that there exists a non-critical
contribution to the density anomaly, which presumably modulates the
critical anomalies.
\bibitem{TDL} We note that if the correlation length of the pseudo
critical freezing transition is finite, then in the thermodynamic
limit, the form of $p(\rho)$ at the transition point must be Gaussian.
The double peaked structure we observe therefore indicates that for the
system sizes studied in this work, the limit $L\gg\xi$ has not been
reached.
\end{thebibliography}
\end{document}